\documentclass[sn-nature]{sn-jnl}

\usepackage{graphicx}%
\usepackage{multirow}%
\usepackage{amsmath,amssymb,amsfonts}%
\usepackage{amsthm}%
\usepackage{mathrsfs}%
\usepackage[title]{appendix}%
\usepackage{xcolor}%
\usepackage{textcomp}%
\usepackage{manyfoot}%
\usepackage{booktabs}%
\usepackage{algorithm}%
\usepackage{algorithmicx}%
\usepackage{algpseudocode}%
\usepackage{listings}%
\usepackage{siunitx} 
\usepackage{fontenc,amssymb}
\usepackage{lipsum}
\usepackage{chemformula}
\usepackage[normalem]{ulem}
\usepackage{multicol}

\newcommand{\CI}{\hbox{{\rm C}{\sc \,i}}}

\newcommand{\SIV}{\hbox{{\rm S}{\sc \,iv}}}
\newcommand{\HeI}{\hbox{{\rm He}{\sc \,i}}}

\newcommand{\FeII}{\hbox{{\rm Fe}{\sc \,ii}}}
\newcommand{\FeIII}{\hbox{{\rm Fe}{\sc \,iii}}}

\newcommand{\SII}{\hbox{{\rm S}{\sc \,ii}}}
\newcommand{\SIII}{\hbox{{\rm S}{\sc \,iii}}}

\newcommand{\NiII}{\hbox{{\rm Ni}{\sc \,ii}}}

\newcommand{\OI}{\hbox{{\rm O}{\sc \,i}}}

\newcommand{\MgII}{\hbox{{\rm Mg}{\sc \,ii}}}

\newcommand{\HI}{\hbox{{\rm H}{\sc \,i}}}

\newcommand{\Ha}{\hbox{{\rm H}$\alpha$}}

\newcommand{\ArII}{\hbox{{\rm Ar}{\sc \,ii}}}
\newcommand{\ArIII}{\hbox{{\rm Ar}{\sc \,iii}}}
\newcommand{\NeII}{\hbox{{\rm Ne}{\sc \,ii}}}
\newcommand{\NeIII}{\hbox{{\rm Ne}{\sc \,iii}}}
\newcommand{\SiIII}{\hbox{{\rm Si}{\sc \,iii}}}

\newcommand{\mpy}{\hbox{M$_{\odot}$~yr$^{-1}$}}

\newcommand{\msun}{\hbox{M$_{\odot}$}}
\newcommand{\cmsq}{\hbox{cm$^{-2}$}}

\newcommand{\um}{\textmu m}
\newcommand{\kms}{\hbox{km~s$^{-1}$}}

\newcommand{\arcsec}{\hbox{$^{\prime\prime}$}}
\theoremstyle{thmstyleone}%
\theoremstyle{thmstyletwo}%

\theoremstyle{thmstylethree}%

\begin{document}
\title[]{Solar C$/$O ratio in planet-forming gas at 1 au in a highly irradiated disk.}


\author*[1]{\fnm{Ilane} \sur{Schroetter}}
\author[1]{\fnm{Olivier} \sur{Berné}}
\author[2]{\fnm{Emeric} \sur{Bron}}
\author[3]{\fnm{Felipe} \sur{Alarcon}}
\author[1]{\fnm{Paul} \sur{Amiot}}
\author[3]{\fnm{Edwin A.} \sur{Bergin}}
\author[4]{\fnm{Christiaan} \sur{Boersma}}
\author[5, 6, 7]{\fnm{Jan} \sur{Cami}}
\author[8]{\fnm{Gavin A. L.} \sur{Coleman}}
\author[9]{\fnm{Emmanuel} \sur{Dartois}}
\author[10]{\fnm{Asuncion} \sur{Fuente}}
\author[11]{\fnm{Javier R.} \sur{Goicoechea}}
\author[12]{\fnm{Emilie} \sur{Habart}}
\author[8]{\fnm{Thomas J.} \sur{Haworth}}
\author[1]{\fnm{Christine} \sur{Joblin}}
\author[2]{\fnm{Franck} \sur{Le Petit}}
\author[13]{\fnm{Takashi} \sur{Onaka}}
\author[5, 6, 7]{\fnm{Els} \sur{Peeters}}
\author[14]{\fnm{Markus} \sur{Rölling}}
\author[15, 16]{\fnm{Alexander G. G. M.} \sur{Tielens}}
\author[12]{\fnm{Marion} \sur{Zannese}}


\affil*[1]{\orgname{Institut de Recherche en Astrophysique et Planétologie, Université de Toulouse, CNRS, CNES}, \orgaddress{\street{ 9 Av. du colonel Roche}, \city{Toulouse Cedex 4}, \postcode{31028}, \country{France}}}
\affil[2]{Laboratoire d'Etudes du Rayonnement et de la Matière, Observatoire de Paris, Université Paris Science et Lettres, Centre National de la Recherche Scientifique, Sorbonne Universit\'es, F-92190 Meudon, France}

\affil[3]{Department of Astronomy, University of Michigan, Ann Arbor, MI 48109, USA}
\affil[4]{NASA Ames Research Center, Moffett Field, CA 94035-1000, USA}
\affil[5]{Department of Physics \& Astronomy, The University of Western Ontario, London ON N6A 3K7, Canada}
\affil[6]{Institute for Earth and Space Exploration, The University of Western Ontario, London ON N6A 3K7, Canada}
\affil[7]{Carl Sagan Center, Search for ExtraTerrestrial Intelligence  Institute, Mountain View, CA 94043, USA}
\affil[8]{Astronomy Unit, School of Physics and Astronomy, Queen Mary University of London, London E1 4NS, UK}
\affil[9]{Institut des Sciences Mol\'eculaires d'Orsay, Universit\'e Paris-Saclay, Centre National de la Recherche Scientifique, 91405 Orsay, France}
\affil[10]{Centro de Astrobiología (CAB), CSIC-INTA, Ctra. de Ajalvir, km 4, Torrejón de Ardoz, 28850 Madrid, Spain}
\affil[11]{Instituto de F\'{\i}sica Fundamental  (Consejo Superior de Investigacion Cientifica), 28006, Madrid, Spain}
\affil[12]{Institut d'Astrophysique Spatiale, Universit\'e Paris-Saclay, Centre National de la Recherche Scientifique, 91405 Orsay, France}
\affil[13]{Department of Astronomy, Graduate School of Science, The University of Tokyo, Tokyo 113-0033, Japan}
\affil[14]{Physikalischer Verein – Gesellschaft für Bildung und Wissenschaft, Robert-Mayer-Str. 2, 60325 Frankfurt, Germany}
\affil[15]{Astronomy Department, University of Maryland, College Park, MD 20742, USA}
\affil[16]{Leiden Observatory, Leiden University, P.O. Box 9513, 2300 RA Leiden, The Netherlands}


\abstract{
The chemical composition of exoplanets is thought to be influenced by the composition of the disks in which they form. JWST observations have unveiled a variety of species in numerous nearby disks, showing significant variations in the C/O abundance ratio. However, little is known about the composition and C/O ratio of disks around young stars in clusters exposed to strong ultraviolet (UV) radiation from nearby massive stars, which are representative of the environments where most planetary systems form, including ours.
We present JWST spectroscopy of d203-504, 
a young 0.7 $\rm M_{\odot}$ star {in the Orion Nebula} with a 30 au disk irradiated by nearby massive stars. These observations reveal spectroscopic signatures of CO, \ch{H2O}, \ch{CH3+}, and PAHs. { Water and CO are detected in absorption} in the inner disk ($r\lesssim 1$ au), where the estimated gas-phase C/O ratio is 0.48, consistent with the Solar value and that of the Orion Nebula. 
In contrast, \ch{CH3+} and PAHs are found in the extended surface layers of the disk.
These results suggest that gas in the inner disk is chemically shielded from UV radiation while the surface layers of the disk experience UV-induced chemistry, potentially depleting their carbon content.
}
\maketitle

Understanding how exoplanets form and acquire their chemical composition is a central question in contemporary astrophysics. These processes take place in protoplanetary disks of gas and dust around young stars, where complex physical and chemical interactions set the initial conditions for planet formation and potential habitability. 

JWST \cite{gardner2023james} is currently transforming our knowledge of the chemical inventory in the planet-forming regions of protoplanetary disks around {nearby }isolated low-mass stars \cite{kamp2023chemical, vanDishoeck_chemistry_jwst_2023, henning2024minds}. Observations have revealed a diverse and widespread presence of mid-infrared emission from neutral species, including \ch{H2O}, \ch{HCN}, \ch{CH4}, and \ch{C2H2}, \cite{grant2023minds, Tabone_disk_chemistry_2023, kospal2023jwst, Perotti_water_23, pontoppidan2024high, colmenares2024jwst}.
Whether {the chemical composition of those disks} 
is dominated by \ch{H2O} or hydrocarbon emissions is thought to be driven by the time evolution of the C/O ratio in the inner disk, which is a key parameter for the chemistry of exoplanets \cite{madhusudhan2012c}.
The prevailing scenario derived from JWST (and earlier Spitzer) observations, supported by models \citep{mah2023close}, suggests an evolutionary sequence: initially, icy pebble drift delivers oxygen to the inner disk through the sublimation of \ch{H2O} at the snowline, reducing the C/O ratio relative to the initial value of the parent {molecular cloud}. 
{Subsequently,
the C/O ratio is expected to increase again. This can be explained 
either by carbon-rich grains and polycyclic aromatic hydrocarbons (PAHs) drifting inward and re-enriching the gas in carbon at the soot-line \citep{vanDishoeck_chemistry_jwst_2023, colmenares2024jwst};
or through viscous transport of carbon-rich gas towards the inner disk \citep{mah2023close}}. The timing of this sequence depends on the {dust drift timescale and} rate of viscous evolution in the disk, {the latter }which progresses faster around low-mass stars compared to solar-type stars \citep{mah2023close}. 

While studies of isolated disks are key to understanding disk chemistry and its link to exoplanet compositions, most low-mass stars form in stellar {group{s}}
that include massive stars \citep{lada2003embedded}. These massive stars ($M_{\rm star} \ge 8 M_{\odot}$) emit strong ultraviolet (UV) radiation, including ionizing extreme ultraviolet (EUV; $E > 13.6$ eV) and dissociating far ultraviolet (FUV; $6 , \mathrm{eV} < E < 13.6 , \mathrm{eV}$) photons. As a result, the majority of planets likely form under conditions influenced by pervasive UV radiation, which can significantly affect star and planet formation \citep{winter2022external}. This scenario is relevant to the Solar System itself, which likely formed in the vicinity of massive stars \citep{adams2010birth, bergin2023interstellar, desch2024sun}. Despite the expected importance of UV radiation in typical star-forming environments, relatively few studies have focused on its impact on the chemical inventory during planet formation \citep{Terada_2012_ONC, Boyden_2023_chemical_modeling_ONC, diaz2024chemistry, Ballering_2025_water_nircam}.

\begin{figure}[h!]
\centering
\includegraphics[width=1\textwidth]{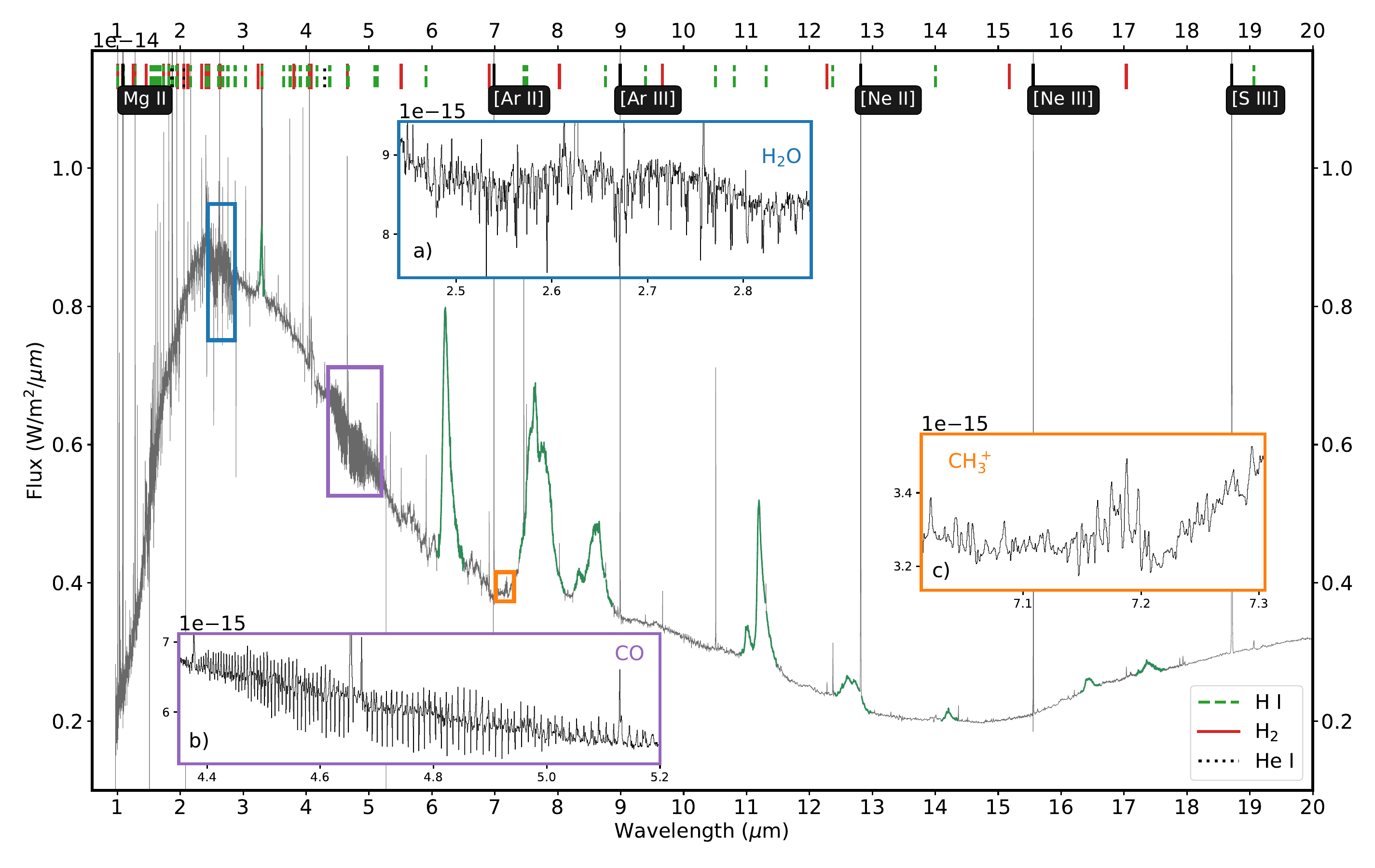}
\caption{ { JWST NIRSpec/IFU and MIRI/MRS combined spectrum of the d203-504 proplyd}. The vertical lines indicate the position of HI (dashed), He I (dotted) and \ch{H2} (continuous) emission lines. Fine structure lines of atoms are labeled directly on the figure.
Three close-up views are displayed in framed insets: water absorption at 2.7~\um\ in blue ({ a)}), CO absorption and emission at 4.7~\um\ in purple ({ b)}) and \ch{CH3+} emission at 7.2~\um\ in orange ({ c)}). Emission bands from PAHs are highlighted in green. }
\label{spectrum}
\end{figure}

Theoretical models have long suggested that strong external UV radiation can significantly alter disk chemistry \citep{walsh2013molecular, champion2017herschel, Gross_2025_chemistry_model, Keyte_2025_chemical}. This has been confirmed by recent JWST observations of the protoplanetary disk d203-506 in the Orion Nebula Cluster (ONC), which is exposed to a strong external FUV radiation field of $G_0 = (2-4)\times10^{4}$ \cite{habart2023pdrs4all} ($G_0=1$ corresponds to the standard interstellar radiation field defined by \cite{Habing68}) . 
Vibrational emission of FUV-pumped \ch{H2} detected with JWST-NIRSpec \cite{BerneO_2024_science} highlights the significant role of FUV photons in gas heating, as seen in photodissociation regions (PDRs) with JWST (see e.g., \cite{peeters2023pdrs4all}). This warm \ch{H2} drives the formation of reactive species like \ch{CH^+}, leading to the production of \ch{CH3+}, a key molecular ion indicative of active organic gas-phase chemistry triggered by FUV radiation \citep{berne2023formation}.
Additionally, the detection of highly rotationally excited \ch{OH} implies efficient \ch{H2O} destruction by UV photons, consistent with the absence of \ch{H2O} in the spectrum of d203-506 \cite{Zannese_24_oh}. Emission from Polycyclic Aromatic Hydrocarbons (PAHs), observed in this disk's JWST spectrum \cite{BerneO_2024_science}, further supports the influence of FUV-driven chemistry, and contrasts with isolated disks, where PAHs are rarely observed \citep{Geers_pah_detection_2007, vanDishoeck_chemistry_jwst_2023}. These findings suggest that UV radiation plays a crucial role in shaping disk chemistry and, consequently, the chemical inheritance passed to forming planets.

However, several limitations remain. The current understanding is based on a single, edge-on disk (d203-506), restricting observations to the extended disk surface and preventing direct assessment of inner disk chemistry, including the C/O ratio. Observations of the XUE1 disk in the massive star-forming region NGC 6357 \citep{xue_ramirez_tannus_2023} show molecular emission lines resembling those of isolated disks, including \ch{H2O}, suggesting limited UV influence.
\cite{diaz2024chemistry} have also shown that {there is no clear signpost of externally UV-driven chemistry in the ALMA spectra of irradiated disks in the outskirts of the Orion Nebula, where the FUV field is $\lesssim 10^3$ $G_0$.}
Recent models also indicate that the inner disk's C/O ratio may remain unaffected by external UV-driven photoevaporation \cite{ndugu2024external}. These discrepancies suggest that the effects of UV radiation on disk chemistry are complex, potentially varying between inner and outer disk regions. To address these uncertainties, we turn to another externally UV-irradiated protoplanetary disk in Orion, d203-504, observed with JWST using both NIRSpec and MIRI instruments {(program id $\#1288$)}.

\section*{Results}\label{results}
The d203-504 {(2MASS J05352025-0525037)} protoplanetary disk is located in the Orion Nebula, a few arcseconds southeast of the ionization front of the Orion Bar \cite{ODellWen94ONC} and $\approx 1.7\arcsec$ to the north of the protoplanetary disk d203-506 \citep{bally2000disks} (see Figure~\ref{fig:rgb}). 
Based on the geometry of the ionization front, d203-504 is predominantly irradiated by the B star $\theta^2$ Ori A \citep{bally2000disks, odell_2017_stars_orion}, which is situated at a projected distance of 0.077~pc (40\arcsec) to the northeast. 
{ Taking the projected separation from the UV sources}, Haworth et al. \cite{haworth2023vlt} estimate the {upper limit of the} FUV radiation field received by d203-504 is $G_0=8\times10^4$ in units of the Habing \cite{Habing68} field. 
Using recent observation with the MUSE instrument on the VLT, the authors also derived the velocity of a jet seen in [\FeII] emission, associated with the Herbig Haro object HH 519 \cite{bally2000disks}, being typically $130~\kms$. The central star is estimated to be a type K5.5 with a mass of 0.72~\msun\ \citep{Fang2021_MUSE_HR_orion, Aru2024_museproplyds}. Parameters for d203-504 are summarized in Table~\ref{table:system_parameters}. 

Using the MUSE data presented in \cite{haworth2023vlt}, we derive a disk size ($r_d = 31$ au) and inclination ($i \geq54^\circ$) as shown in the Methods. d203-504 has also been observed with ALMA (PI J. Champion) together with d203-506 \cite{BerneO_2024_science}. d203-504 is detected only in the continuum at 344 GHz (Figure~\ref{fig:alma}), from which we derive an estimated disk mass of $M_{d}=1.3-5.6~{\rm M}_{\rm Jup}$ (see Methods).

\begin{table}[h!]
\centering
\caption{Summary of d203-504 disk main properties}
\label{table:system_parameters}
\begin{tabular}{llcc}
\hline
Parameter  &  &  Value(s) & Origin  \\
\hline
$D$                 & Distance & 390 pc & 3 \\
$T_\star$           & Stellar type, temp. & K5.5, 4160~K  & 2 \\
$r_\star$           & Stellar radius & 0.7 R$_{\odot}$  & 2 \\
$M_\star$           & Central star mass & 0.72~\msun\ & 2 \\
$V_{\rm sys}$       & Systemic velocity & 25 \kms & 1 \\
$G_0$               & Radiation field & $\lesssim$8$\times 10^4$ & 1 \\
$V_{\rm Jet}$       & Jet velocity & 130~\kms & 1 \\
$\dot M_{\rm PE}$   & Photoevaporation rate & $2.5 - 7.9 \times 10^{-7}$\mpy & 1,2,4  \\
$\dot M_{\rm acc}$  & Accretion rate & 8.2$\times 10^{-9}$\mpy & 4  \\
$r_d$               & Disk radius & 31 au (0.08\arcsec) & 4  \\
$E_d$               & Disk thickness & 46 au (0.12\arcsec) & 4  \\
$i$                 & Disk inclination & $\gtrsim 54^\circ$ & 4  \\
$F_{\rm 344}$       & Dust flux at 344 GHz & 7.4 mJy & 4 \\
$M_{\rm disk}$      & Disk mass & 1.3--5.6 M$_{\rm Jup}$ & 4 \\
\hline
\end{tabular}
{1: \cite{haworth2023vlt}, 2: \cite{Aru2024_museproplyds}, 3: \cite{apellaniz2022villafranca}, 4: this work.}
\end{table}

d203-504 was observed as part of the JWST early release science program PDRs4All \citep{pdrs4all2022} with NIRCam \citep{rieke2023performance}, MIRI MRS \cite{wells_miri_mrs2015} and NIRSpec IFU \citep{boker_nirspec_ifu_2022}. Those observations are presented in \cite{habart2023pdrs4all}, \cite{van2024pdrs4all, chown2023pdrs4all}, and \cite{peeters2023pdrs4all}, respectively.
In Figure~\ref{spectrum}, we present the background subtracted combined NIRSpec-MIRI spectrum of this source (see Methods).
This spectrum is characterized by a strong near-infrared continuum. This continuum is well fitted by a 1180~K blackbody, with a luminosity $L_{\rm NIR}=0.13~{\rm L}_{\odot}$ (see Methods). This near-infrared excess is  most likely related to viscous heating linked to accretion. The derived luminosity  corresponds to an accretion rate of $\dot{M}_{\rm acc}=8.2\times10^{-9}$ \mpy, typical for young T-Tauri stars \citep{rigliaco2013accretionttauri}. 
Emission from the characteristic bands of PAHs are well detected at 3.3, 6.2, 7.7, 8.6, 11.2, 12.7~\um\ as well as several other fainter bands, also seen in the Orion Bar with JWST \cite{chown2023pdrs4all}.
We note that this PAH emission clearly arises from the disk, as there is a strong excess in emission in all PAH bands with respect to the background nebula (see Figure~\ref{fig:pah_on_vs_off}). A broad feature between 7 and 13~\um, corresponding to the emission of hot silicate grains, is also present (this feature is best seen in the spectrum presented in Figure~\ref{fig:spectrum_mjy}).
In addition to these dust-related features, numerous atomic and molecular lines are present (Figure~\ref{spectrum}). These include fine structure lines from atoms and atomic ions, recombination lines of H and He, and molecular hydrogen lines. The full list of lines and assignments is provided in Tables~\ref{table:line_intensities} and~\ref{table:line_intensities_2}.
We also detect spectroscopic signatures from several molecules, with NIRSpec: \ch{H2O} $v=1 \rightarrow 0, \nu_1, \nu_2$ symmetrical and asymmetrical stretching, in absorption, between 2.5 and 2.9~\um{ (to our knowledge, this is the first time these near IR \ch{H2O} bands are detected in protoplanetary disks as they are very difficult to observe from the ground)};  CO $v=1 \rightarrow 0$, $v=2\rightarrow1$ in emission and absorption around 4.4 -- 5~\um. In addition, we tentatively detect several OH and \ch{CH+} lines (see Figure~\ref{fig:oh_chp}).  
With MIRI, we detect the \ch{CH3+} 
$\nu_2^+$ out-of-plane and $\nu_4^+$ in-plane dyad modes \citep{berne2023formation, changala2023astronomical} near 7.2~\um. 
Finally, we note that \ch{HCN}, \ch{C2H2} or \ch{CO2} emission signatures that are commonly detected in isolated disks 
are not seen in d203-504. 
The molecular detections, and non-detections, reported here bring an additional piece to the puzzle of chemistry in externally irradiated disks. 
In the following, we address the origin of the spectral signatures seen in emission and in absorption in d203-504.
\\

\begin{figure}[h!]
\centering
\includegraphics[width=1.\textwidth]{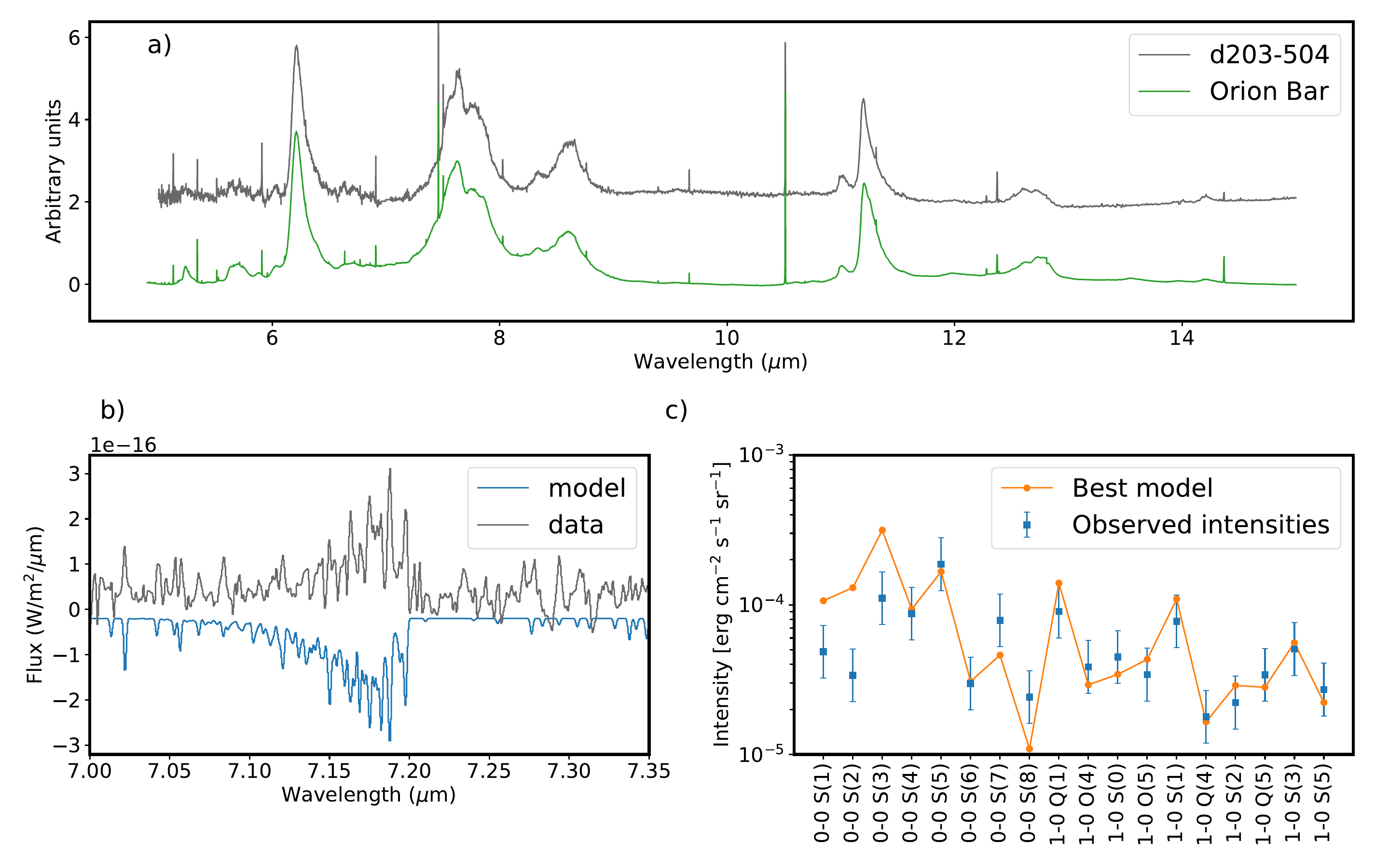}
\caption{{ Analysis of PAH, \ch{CH3+} and \ch{H2} emissions in d203-504}.
{ a)} Continuum subtracted PAH spectrum ($F_{\rm PAH}(\lambda)$, see Methods) in arbitrary units, for d203-504 (gray) and the Orion Bar PDR (green). 
{ b)} Continuum subtracted spectrum of d203-504 ($F_{\rm obs}(\lambda) - F_{\rm cont}(\lambda)$, see Methods) in the spectral region of  \ch{CH3+} emission, in dark gray and the best fit model in blue.
{ c)} Comparison between the observed (blue) and modeled (orange) \ch{H2} line intensities using the Meudon PDR model. Error bars represent a total uncertainty of 50\% used in the minimization to obtain the best fit model. The instrumental uncertainties are smaller than the point size and given in Table~\ref{table:model_properties}.
Notations of \ch{H2} transitions on the x-axis are abbreviated. The corresponding quantum levels are given in Table~\ref{table:h2_em}.}
\label{fig:pah_ch3p}
\end{figure}

We {first} focus on the strongest detections observed in emission arising from the UV-irradiated neutral layer at the disk surface, that is \ch{CH3+}, \ch{H2}, and PAHs. CO is also detected in emission, but the broad emission profiles suggest that this emanates from the outflow or wind rather than the disk itself (see Methods). 
\ch{CH3+} emission has recently been detected in the d203-506 externally irradiated disk \cite{berne2023formation},
the isolated disk TW Hya \citep{henning2024minds}, and the Orion Bar PDR \cite{zannese_2024_ch3p}. 
This cornerstone cation is attested to be formed through 
 a chemical reaction sequence initiated by the reaction of \ch{C+} and vibrationally excited \ch{H2}
by FUV radiation \cite{zannese_2024_ch3p}. \ch{CH3+} is thus a chemical marker of FUV driven chemistry. 
We model the emission of \ch{CH3+} (see Methods) and find an excitation temperature of $T_{\rm ex} = 1000\pm 100$ K, slightly higher but close to that found in d203-506 ($T_{\rm ex} = 660 \pm80$ K by \cite{changala2023astronomical}) and for the Orion Bar (1250$\pm250$~K, \cite{zannese_2024_ch3p}). 
The column density  of \ch{CH3+} is $N_{\rm CH_3^+}=4\times10^{13}$~\cmsq (see Methods). 
This is higher than the value reported by \cite{changala2023astronomical} for d203-506 ($N_{\rm CH_3^+}=5\times10^{11}$~\cmsq), however these latter authors considered a beam averaged value whereas we use 
the physical size of the disk (see Methods).  
In Figure~\ref{fig:pah_ch3p}, we present the extracted PAH spectrum  of d203-504 (see Methods) and compare it 
to the Orion Bar atomic PDR zone \cite{chown2023pdrs4all}.
The two spectra are very similar, indicating that PAHs in d203-504 have a 
composition close to that of the Orion Nebula. 
Following the analysis of \cite{BerneO_2024_science}, we use the observed \ch{H2} fluxes (Table~\ref{table:h2_em}) in the detected ro-vibrational lines coupled with the Meudon PDR code \cite{LePetitF_06} to derive the physical conditions in the PDR on the disk surface (see Methods). 
The best fit model is shown in Figure~\ref{fig:pah_ch3p}c, and yields a gas temperature $T_{\rm gas} = 1240$~K, a hydrogen number density $n_{\rm H}=3.1\times10^7$ cm$^{-3}$, and a photoevaporation mass-loss rate $\dot{M}_{\rm PE}=7.9\times10^{-7}$~\mpy, slightly higher than earlier indirect estimates {under the assumption of photoionization equilibrium}
($\dot{M}_{\rm PE}= 2.5 \times 10^{-7}$~\mpy, \cite{Aru2024_museproplyds} and $\dot{M}_{\rm PE}= 3.1 \times 10^{-7}$~\mpy). 
Overall, \ch{CH3+}, \ch{H2}, and PAH emissions are clear indications of active photochemistry taking place in the $n_{\rm H}\sim 10^{7}$ cm$^{-3}$ PDR formed at the surface of the disk by irradiation from external FUV photons.  
\\

\begin{figure}[h!]
\centering
\includegraphics[width=1.0\textwidth]{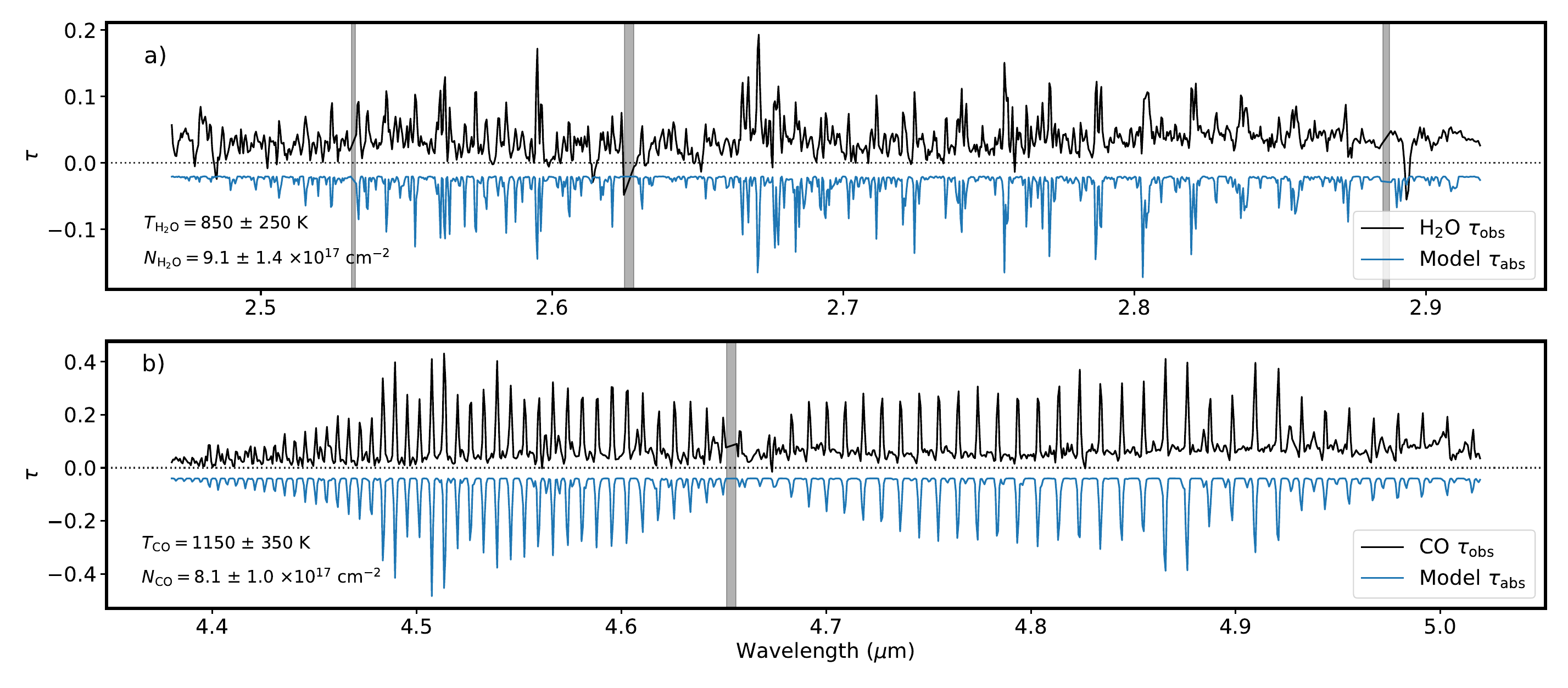}
\caption{{ Optical depth spectra of water and carbon monoxide absorption in the inner disk of d203-504.} { a)} Water absorption optical depth spectrum derived from the observations ($\tau_{\rm obs}$, see Methods), in black, and best fit model ($\tau_{\rm abs}$) shown in negative, in blue. 
{ b)} Carbon monoxide absorption optical depth spectrum derived from the observations, and best fit model in blue.
The gray areas correspond to regions of contamination by strong atomic lines where the data has been {excluded from the fit}.  
Column densities and temperatures obtained with the best fit models are indicated.}
\label{fig:models}
\end{figure}

We {then} model the absorption in the $v=1 \rightarrow 0$ and $v=2 \rightarrow 1$ vibrational {bands} of \ch{CO}, as well as the $v=1 \rightarrow 0$, $\nu_1$, and $\nu_2$ bands of \ch{H2O}, observed against the bright infrared continuum, using a simple slab model at local thermodynamical equilibrium (see Methods). 
The best-fit models, shown in Figure~\ref{fig:models}, yield excitation temperatures of $T_{\rm H_2O} = 850 \pm 250$~K and $T_{\rm CO} = 1150 \pm 350$~K, with column densities of $N_{\rm H_2O} = (9 \pm 1) \times 10^{17}$~cm$^{-2}$ and $N_{\rm CO} = (8 \pm 1) \times 10^{17}$~cm$^{-2}$. 
The high temperatures ($T_{\rm ex} \sim 1000$~K)  combined with the absorption seen against the hot ($T_{\rm NIR}=1180$~K) inner disk emission at a similar temperature indicate that \ch{H2O} and \ch{CO} originate in the inner disk. { In addition, we show in the  Methods, and Figure~\ref{fig:water} that the absorption of water in the $\nu_2$ bending mode supports this interpretation as well.}
{ The column densities derived for \ch{H2O} and \ch{CO} are also compatible with those found in the inner disks of isolated young stars (e.g. \cite{SalykC_08_water_CO, BanzattiA_2022_scanning})}. These molecules must be slightly cooler than the dust to produce absorption. The hot dust continuum is emitted from a region with an outer radius of $r_{\rm out} = 0.047$~au (Methods).
Assuming a radial temperature profile for the dust, $T_{d}(r) = T_{\rm NIR}(r/r_{\rm out})^{-3/4}$ \citep{facchini2017different}, and a water freeze-out temperature of 150~K implies that the {water} snowline is located at $r_{\rm snow} = 0.73$~au. Therefore, the observed water absorption arises from within this radius.
{ The inner disk column densities of \ch{H2O} and \ch{CO} are derived from absorption, and thus do not depend on an emitting surface area (Eq. \ref{eq:model_abs}). The absence of other spectral signatures in absorption, indicates these molecules are likely the main carriers of carbon and oxygen.
In addition, the excitation temperature for the two species are compatible within errors, suggesting they are located in the same region. Therefore, we can derive an estimate of the C/O ratio in the inner disk
using C/O$= N_{\rm CO}/(N_{\rm CO}+N_{\rm H_2O}) = 0.47 \pm 0.07$,} consistent with the Solar value of $0.51 \pm 0.06$ \citep{bergemann2021solar}, and the value derived for the Orion Nebula, $0.52 \pm 0.18$ \citep{cartledge2001space}.

\section*{Discussion}\label{discussion}

\begin{figure}[h!]
\centering
\includegraphics[width=1.0\textwidth]{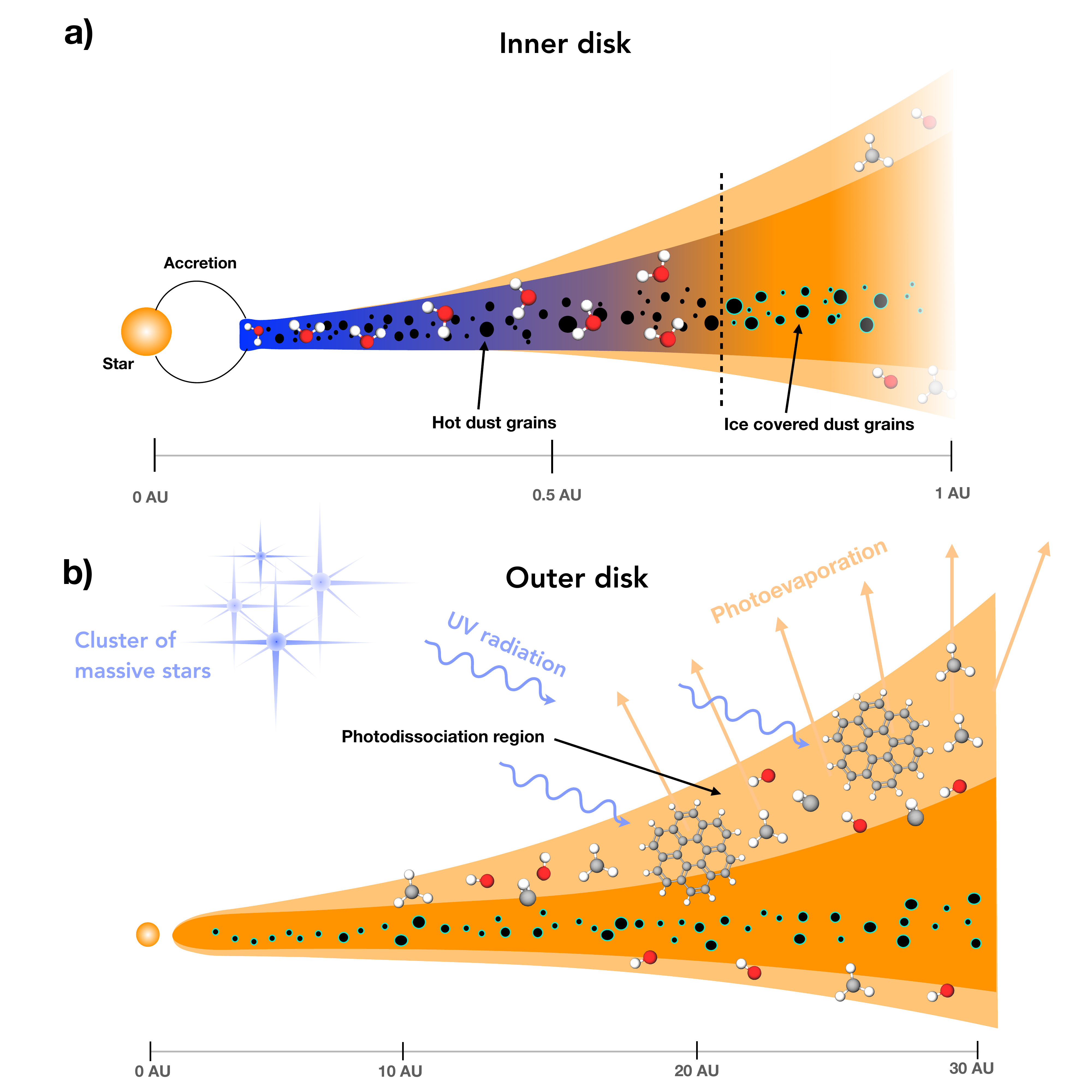}
\caption{{ Chemical structure in the solar-type d203-504 Proplyd.}
{ a)} Sketch of the inner disk ($\lesssim 1$~au), composed of hot dust grains and water vapor heated by accretion processes. 
This heating sets the position of the snowline, that is, the radius at which the water vapor can freeze-out on dust grains, at $r\sim 0.7$~au (vertical dashed line). 
{ b)}  Sketch of the outer disk ($\gtrsim 1$~au). The surface layers host a photodissociation region with photo-chemistry involving \ch{CH+}, \ch{CH3+}, OH and PAHs driven by the FUV photons from nearby massive stars. 
This PDR triggers mass-loss by photoevaporation.}
\label{scheme}
\end{figure}

The analysis of the molecular emission and absorption observed with JWST in the spectrum of d203-504 points to the existence of two distinct chemical regions in the disk, as depicted in Figure~\ref{scheme}. \ch{H2O} and CO observed absorptions originate from the inner disk ($r\lesssim 1$~au), while \ch{CH3+}, \ch{H2}, and PAH emissions originate from the PDR at the base of the FUV driven photoevaporation flow extending to the outer disk ($r\gtrsim 1$~au). 

In the inner disk, within the snowline, water is present in the gas phase. The presence of large column densities of gas phase \ch{H2O} in the inner disk, together with the absence of OH emission, implies that water is not photodissociated by UV, and thus shielded from UV radiation in these dense regions of the disk \cite{bethell2009formation, duval2022water}. This indicates that the inner disk chemistry (in the sense of the chemical reactions that occur) is not directly affected by the external FUV radiation, and instead driven by ice sublimation at the surface of viscously heated dust, as observed in isolated disks \cite{vanDishoeck_chemistry_jwst_2023, kospal2023jwst, BanzattiA_2023_kinematics, henning2024minds}. 
The detection of water in absorption in d203-504, as opposed to its absence in d203-506, likely arises from d203-504's geometry having a favorable vewing angle. In contrast, d203-506 is almost perfectly edge-on, causing the disk to obscure the warm dust emission and any associated water absorption from the inner regions.

The C/O ratio (0.48) and absence of hydrocarbons (e.g. \ch{C2H2}, HCN) in the inner disk indicates the gas is oxygen rich, with a value close to Solar and to the ISM of the Orion Nebula. This value can be interpreted in the context of the recent models of 
\cite{ndugu2024external}, which predict the evolution of the C/O ratio in the inner regions of an externally irradiated disk. According to these models, a Solar C/O ratio in the inner disk is compatible only with two cases: either a very young ($<$1 Myr) disk that has not evolved viscously yet, or a more evolved disk { ($\gtrsim$ 2 Myrs) where {some} carbon re-enrichment of the inner disk has already occurred}.
{ The first scenario is compatible with the expected young age of the Orion Nebula Cluster
($\sim 1$Myr \cite{Hillenbrand_97_orion_stellar}) and the fact that  highly irradiated disks like d203-504 are unlikely to live more than a million years. With a photoevaporation mass-loss rate $\dot{M}_{\rm PE}\sim 3\times10^{-7}$\mpy and a disk mass $M_{\rm d}<6$M$_{\rm Jup}$, the disk depletion time scale is only 20,000 years. This scenario is also compatible with the presence of PAHs indicating recent inheritance from the ISM (whereas PAHs are rarely found in more evolved isolated disks \cite{Geers_pah_detection_2007}).  
The second scenario implies that the disk has been shielded from external UV radiation at an earlier point in time, and also that the Orion Nebula is somewhat older (a few Myrs, \cite{fajrin2025low}).}
{In any case the C/O ratio in the inner disk of d203-504 is consistent with that of a young, isolated, viscously evolving disk around a Solar-type star.} 

\begin{figure}[ht!]
  \centering
  \includegraphics[width=12cm]{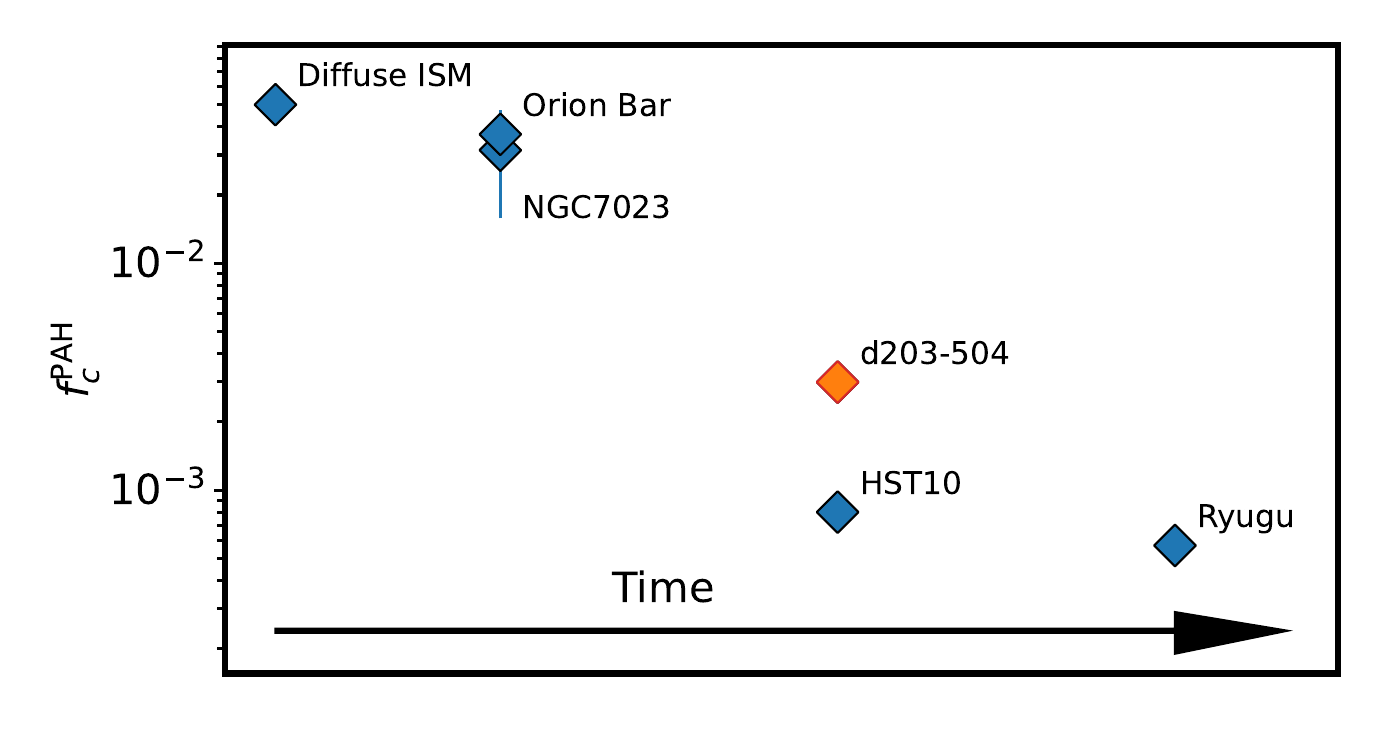}
  \caption{Fraction of the available cosmic carbon abundance locked in PAHs. Those fractions are measured at different times of the stellar evolution (from older to more recent, left to right). The Orion Bar and NGC 7023 are photodissociation region in star-forming regions. HST10 is a proplyd in Orion, situated closer to the T{r}apezium cluster {with respect} to d203-504 with $G_0\sim2\times10^5$. Ryugu is a type C asteroid {and is used as an approximation for the Solar System}. See Methods for details on the derivation of PAH abundances presented in this figure.}
  \label{fig:fcpah}
\end{figure}

The upper layers of the d203-504 disk, extending to its outer edge ($r_{\rm d} = 31$~au) and observed in emission with JWST, are strongly influenced by UV radiation. In these regions, a PDR is formed, where chemistry is driven by ionized carbon and FUV–pumped \ch{H2}. This results in the formation of reactive species such as \ch{CH+} and \ch{CH3+} \citep{berne2023formation, GoicoecheaJ_PDRs4All_jwst_alma_2024}, while water is photodissociated into OH radicals \citep{Zannese_24_oh}.
PAHs are also present in this PDR, exhibiting emission properties similar to those observed in the Orion Bar (Figure~\ref{fig:pah_ch3p}). This similarity suggests a common origin, with PAHs in d203-504 either inherited from the parent nebula during disk formation—potentially locked in ices or larger grains and later released by FUV-driven photo-chemical processing \cite{berne_analysis_2007, bouwman2011photochemistry}—or formed in situ via gas-phase reactions in the PDR \citep{Cox_2016_pah_nebula}. 
PAHs are simultaneously subject to UV-induced processing (e.g., removal of labile aliphatic \ch{CH} bonds \cite{marciniak2021photodissociation}) and destruction via photodissociation \cite{montillaud_evolution_2013, Andrews2016} or reactions with oxygen \cite{lee2010solar}.
An important question is how this surface-layer photochemistry influences the disk midplane, where planets form. The observed species in the PDR are embedded in hot gas that escapes the disk’s gravitational potential in a photoevaporative flow. Unless vertical mixing driven by turbulence is particularly efficient \cite{semenov2010chemistry}, these species will be lost in the wind rather than incorporated into the midplane. 
However, in the scenario where vertical mixing is efficient, Lange et al. \cite{Lange_2023_pah_cycle} proposed that PAHs could evolve through a processing cycle during which they mix into the UV depleted layers where they collide, leading to sticking and eventually freeze out on dust grains. Small clusters of PAHs whould desorb while the large ones would remain on the grains. Repetitions of this cycle would lead to depletion of gas-phase PAHs while large clusters would remain on dust grains. 
In Figure~\ref{fig:fcpah}, we compare the fraction of the derived cosmic carbon abundance locked in PAHs in d203-504 ($f_C^{\rm PAH}= 2.9\times 10^{-3}$, see Methods) with that of the ISM and the Solar System. The observed trend suggests a progressive depletion of PAHs over time, consistent with their destruction at the disk surface. This implies that a reduced fraction of PAHs compared to the ISM are incorporated into the solids of planetary systems exposed to strong UV.
PAHs represent a potentially significant carbon reservoir, traditionally thought to contribute to the gas-phase carbon at the ``soot line'' in protoplanetary disks, fueling a rich organic chemistry \cite{kress2010soot, li2021earth, vanDishoeck_chemistry_jwst_2023, colmenares2024jwst}. However, in disks like d203-504, where PAHs are destroyed at the surface, the soot line may be suppressed (which is also compatible with the absence of small hydrocarbons discussed hereabove {and might be an indirect effect of external UV on the inner disk}). This selective loss of PAHs while oxygen-rich grains remain gravitationally bound suggests that externally UV-irradiated disks are likely to remain oxygen-rich. 
The loss of PAHs also reduces UV opacity and photoelectric heating efficiency in the PDR, potentially altering gas heating, chemistry, and photoevaporation dynamics \cite{facchini2016external}. 

To fully understand the impact of UV irradiation on the chemical composition of planet-forming material, future efforts must couple dynamical and photo-chemical models of irradiated disks with extensive observational surveys of externally illuminated disks in various star-forming regions.

\section*{Correspondence}
Correspondance and request for materials should be addressed to Ilane Schroetter, ilane.schroetter@gmail.com.

\newpage

\section*{Methods}\label{methods}

\makeatletter

\noindent
\subsection*{MUSE observations of d203-504}
\label{203-504_muse}

The MUSE observations used in this paper are described in details in \cite{haworth2023vlt}. 
{Left panel of }Figure~\ref{fig:rgb} shows composite image of both d203-506 (lower left) and d203-504 (top right) systems created using those data.  d203-504 shows a clear ionization front in \Ha, faint jets in [\FeII] and the disk in [CI]. 
{Right panel of Figure~\ref{fig:rgb} shows the same field of view but with JWST/NIRCam observations. On this panel, d203-504 is saturated.}
Th{e [CI]} emission line is a tracer of the disk surface \cite{GoicoecheaJ_PDRs4All_jwst_alma_2024, Aru2024_museproplyds, aru2024tell}. 
We thus fit a 2D Gaussian to this emission  and obtain a disk radius $r_{d}\sim $31 au (0.08\arcsec\ at 390~pc){, corresponding to the semi-major axis divided by two,} and a disk thickness 
$E_{\rm d} \sim 46$ au (0.12\arcsec){, corresponding to the semi-minor axis}. In addition, we can obtain an estimate of the lower limit for the disk inclination $i$ {(the inclination convention here is to be 0$^\circ$ for an face-on disk and 90$^\circ$ for edge-on configuration)}, following the method of \cite{BerneO_2024_science}. That is, $i\gtrsim90-\rm{arctan}(\frac{E_{\rm d}}{2r_{\rm d}})$, yielding $i\gtrsim 54^\circ$.
This estimation is in agreement with  the jet being tilted by at least 45$^\circ$ \cite{haworth2023vlt}.

\subsection*{ALMA observations of d203-504}
\label{alma}

The ALMA data and associated data reduction used in this paper was presented in details in \cite{BerneO_science}. In Figure~\ref{fig:alma} we show the 344 GHz ALMA dust continuum emission map of d203-506 and d203-504. 
We extract a flux for d203-504 $F_{344}= 7.4 $~mJy using a 2D Gaussian fit. 
The disk, however, is not resolved in the ALMA observations and thus cannot provide with another estimate of the disk size. Following the method presented in \cite{mann2014alma} (see also \cite{BerneO_science}) 
we derive a disk mass range of 1.3--5.6 M$_{\rm Jup}$ using this flux. 
{This range of values stems from the uncertainties on the value of the $\kappa_\nu$ parameter, the dust grain opacity, and the optical depth of the emission at 344 GHz. This is explained in details in \citep{BerneO_science}.}
{ We also note that d203-504 is present in the {VLA} catalog of \cite{forbrich2016population} and has a peak intensity of 0.1886 mJy {at their reference frequency of 6.1 GHz}. 
However, the 344 GHz ALMA data {has a peak intensity of $\sim 3.0$~mJy and }is peaked at the disk position, not at the ionization front. Thus the ALMA 344 GHz emission must be largely dominated by the thermal dust rather than free-free emission. 
}

\subsection*{JWST observations and data reduction.}
\label{jwst_dr}
Observations were performed with the JWST-NIRSpec and JWST-MIRI-MRS Integral Field Units (IFUs) as part 
of the PDRs4All ERS program \citep{pdrs4all2022}. 
They were reduced using the JWST pipeline version 1.10.2 with calibration 
reference data system (CRDS) context file jwst\_1084.pmap. 
Data cubes are a combination of 9 pointings, forming 
a mosaic spanning across the Orion Bar. The detailed data reduction together with a general 
analysis of line emissions are described in \cite{peeters2023pdrs4all} and \cite{chown2023pdrs4all, vanDishoeck_chemistry_jwst_2023} for NIRSpec-IFU and MIRI MRS, respectively.
To extract the d203-504 spectrum, we first take a circular aperture of 0.40\arcsec\ radius centered on d203-504 coordinates (right ascension (RA)$=$5:35:20.268 and declination (DEC)$=$-05:25:03.992), which we call ``ON''. 
Then, in order to subtract the background emission from the nebula, we extract a spectrum of similar aperture (0.365\arcsec) at coordinates RA=5:35:20.370 and DEC$=$-05:25:04.97 where no obvious object can be detected{(we note that this background spectrum, for consistency, is extracted using the same aperture and position as used by \cite{berne2023formation} and \cite{BerneO_science})}.
We call this position ``OFF''.
The ``ON'' and ``OFF'' spectra are shown in Figure~\ref{fig:pah_on_vs_off}. 
Finally the d203-504 spectrum $F_{\rm obs}(\lambda)$ is obtained by subtracting ``OFF'' from ``ON'' which is shown in Figure~\ref{spectrum}. \\

\noindent
\subsection*{Details on line identification.}
\label{line_id}
Lines from atomic ions and \ch{H2} in the JWST spectrum of d203-504 (Figure~\ref{spectrum}) are identified using the line list provided  by the PDRs4All ERS team{(\url{http://hebergement.u-psud.fr/edartois/jwst_line_list.html})} and benchmarked on the Orion Bar spectroscopic data \citep{peeters2023pdrs4all, vanDishoeck_chemistry_jwst_2023}.
Numerous strong \HI\ emission lines (from $4\rightarrow20$ to $14 \rightarrow 43$) are detected as well as several ionic lines including \MgII, [\ArII], [\ArIII], [\NeII], [\NeIII], [\SII], 
[\SiIII], [\FeII] and [\FeIII].
For \ch{H2}, we detect the $v=0 \rightarrow 0$, $v=1 \rightarrow 0$, $v=1 \rightarrow 1$, $v=2 \rightarrow 0$, $v=2 \rightarrow 1$, $v=3 \rightarrow 2$, $v=3 \rightarrow 1$, $v=4 \rightarrow 3$ and $v=4 \rightarrow 2$ transitions in emission. 
Table~\ref{table:line_intensities} lists the strongest atomic and \ch{H2} lines detected and their corresponding intensities. 
{Each detected emission line is fitted using a Gaussian with a FWHM corresponding to the instrument PSF plus a linear continuum. We integrate the result of the Gaussian fit to derive the intensity}

\noindent
\subsection*{Stellar, dust and PAH continuum emission}
\label{accretion}

We model the continuum flux of d203-504 observed in the NIRSpec and MIRI ranges using the sum of the 
emissions from the star $F_{\star}$,  near-infrared emission from the inner disk, $F_{\rm NIR}$, and 
 mid-infrared emission from the outer disk, $F_{\rm MIR}$:
\begin{equation}
F_ {\rm cont}(\lambda) = F_{\star}(\lambda) + F_{\rm NIR}(\lambda) + F_{\rm MIR}(\lambda),
\label{f_cont}
\end{equation}
where 
\begin{equation}
F_{\star}(\lambda) = \frac{4\pi~r_{\star}^2}{d^2}~B(\lambda, T_{\star})~e^{-\tau_{1\mu m}},
\label{f_star}
\end{equation}
\begin{equation}
F_{\rm NIR}(\lambda) = \frac{\pi~(r_{\rm out}^{2}-r_{\rm in}^2)}{d^2} B(\lambda, T_{\rm NIR})e^{-\tau_{3\mu m}},
\label{f_nir}
\end{equation}
\begin{equation}
F_{\rm MIR}(\lambda) = \sum _iB({\lambda}, T_{\rm MIR}^i)+AIB(\lambda).
\label{f_mir}
\end{equation}
In the above equations, $B$ is the Planck function, $r_{\star}$ is the radius of the star, and the stellar temperature $T_{\star}=4160$~K \cite{Aru2024_museproplyds}. $T_{\rm NIR}$ is the dust temperature in the inner disk. $r_{\rm in}$ is the inner radius of the disk, determined by the sublimation radius of the dust, that is, $r_{\rm in}=0.07 \sqrt{\frac{L_{\star}}{L_{\odot}}}=0.037$ au following \cite{facchini2017different}. $r_{\rm out}$ is the outer radius of the hot inner disk.  
$T_{\rm MIR}^i$ are the mid-infrared dust emission temperatures, and $AIB (\lambda)$ stands for the emission from the aromatic infrared bands of PAHs, 
modelled using a collection of Gaussians as described in detail in \cite{foschino_s_learning_2019}. 
The exponential factors in Eqs.~\ref{f_star} and \ref{f_nir} are a correction for the line of sight extinction, with $\tau_{1\mu m}$ and $\tau_{3\mu m}$ the optical depths at 1 and 3 $\mu$m, respectively. The values of $\tau_{1\mu m}$ and $\tau_{3\mu m}$ are computed using the extinction curve from \cite{draine2011} {with Rv = 5.5}, for a visual { line of sight} extinction { towards the star} $A_{V}=3$ as derived from \cite{Aru2024_museproplyds} for d203-504. 
This yields $\tau_{1\mu m}=4.8$ and $\tau_{3\mu m}=0.25$. 
Figure~\ref{fig:bb} shows the fit of $F_{\rm cont} (\lambda)$ to the observed spectrum $F_{\rm obs} (\lambda)$. 
The fit results provide the values of $T_{\rm NIR}=1180$K, $r_{\rm out}=0.047$ au. The near infrared luminosity is $L_{\rm NIR}=\int{F_{\rm NIR}(\lambda)d\lambda}=0.13~L_{\odot}$. This corresponds to an accretion rate $\dot{M}_{\rm acc}=2 r_{\star}L_{\rm NIR}/(G M_{\star})=8.2\times10^{-9}~{\rm M}_{\odot}/{\rm yr}$.

\noindent
\subsection*{d203-504 disk PAH spectrum}
\label{PAH_spectra}

{ In Figure \ref{fig:pah_on_vs_off} we present the MIRI ON and OFF spectra of d203-504. We fit both continua in these spectra using a collection of black bodies with the PAHTAT \cite{foschino_s_learning_2019} tool. We subtract these continua to the ON and OFF spectra so that they contain only the ON and OFF PAH emissions. We then subtract the OFF PAH emission to the ON PAH emission (Figure \ref{fig:pah_on_vs_off}). This yields the disk PAH emission (Figure \ref{fig:pah_on_vs_off}). The disk PAH emission is presented in Figure~\ref{fig:pah_ch3p} together with the continuum subtracted  Orion Bar spectrum which was obtained with the same method, applied to the template spectrum ``Atomic PDR'' of \cite{chown2023pdrs4all}.}

\noindent
\subsection*{Carbon fraction locked in PAHs}
\label{models}

To derive the abundance of PAHs in d203-504, expressed as the fraction of carbon locked in PAHs $f_C^{\rm PAH}$ we follow the method presented in \cite{vicente2013polycyclic} (Eq. 1), relying on the 8.6~\um\ PAH intensity, $I_{8.6}$:
\begin{equation}
I_{8.6}=\epsilon_{8.6}~f_{\rm C}^{\rm PAH}~\frac{[\rm C]}{[\rm H]}~N_{\rm H}~G_0,
\label{pahem}
\end{equation}
where $f_C^{PAH}$ is the fraction of elemental carbon locked in PAHs, $N_{\rm H}$ is the column density of hydrogen atoms in the line of sight,  $\frac{[C]}{[H]}$ is the abundance of carbon relative to hydrogen atoms ($1.6 \times 10^{-4}$), and $\epsilon_{8.6}$ is the PAH emissivity at 8.6~\um, with $\epsilon_{8.6}=4.5\times10^{-18}$~MJy/sr/cm$^{-2}$/$G_0$ \cite{vicente2013polycyclic}.
Using the extracted PAH spectrum $F_{\rm PAH}(\lambda)$, we derive an intensity $I_{8.6}=F_{\rm PAH}(8.6\mu m)=862$ MJy$/$sr. Using $G_0=8\times10^4$ and $N_{\rm H}=5.5\times 10^{21}$~\cmsq\ as in \cite{vicente2013polycyclic}, we compute a value $f_C^{\rm PAH}=2.9\times 10^{-3}$.

The abundance for HST 10 is taken from \cite{vicente2013polycyclic}, $f_C^{\rm PAH}=8.0\times 10^{-4}$. 
The abundance for Orion Bar is from \cite{TielensA_05} ($f_C^{\rm PAH}=3.7\times 10^{-2}$) and for NGC 7023 ($f_C^{\rm PAH}=3.2\times 10^{-2}$) from \cite{berne_formation_C60_2012}. 

The abundance in the asteroid Ryugu is computed using the result from the Hayabusa mission. 
From \cite{aponte2023pahs} (Figure 2), we obtain a concentration of 70 nmol/g for \ch{C16H10} (Pyerene + Fluoranthene).
From \cite{sabbah2024first} (Figure 6), we obtain a signal of 11576 counts for condensed aromatics with over 30 C atoms ($N_{\rm C} >30$), which are interstellar PAH candidates, whereas the ion signal for \ch{C16H10} is 14535 counts.
Therefore, the abundance of large $N_{\rm C} >30$ PAHs can be estimated by $70\times 11576/14535= 56$ nmol/g. 
The averaged molar mass of large condensed PAHs (\ch{C38H16}) is 472 g/mole. 
Therefore, the mass fraction of large PAHs is estimated to be $56\times 10^{-9}\times 472=2.6\times10^{-5}$. 
The total carbon abundance in Ryugu is 4.6 $\%$ of the total mass \cite{yokoyama2022samples}, thus the fraction of carbon locked in PAHs in Ryugu is derived to be $f_{C}^{\rm PAH}=5.7\times10^{-4}$. 

\subsection*{Fit of rovibrational \ch{H2} lines with the Meudon code}

17 lines of molecular hydrogen are detected towards d203-504. Their intensities are reported in Table~\ref{table:h2_em}.
{ In order to determine the physical conditions of the H$_{\rm 2}$ emitting gas, 
we use the Meudon PDR code \cite{le_petit_model_2006} to fit these intensities.
We used a grid of constant density PDR models produced with the Meudon PDR code (version 7.0, publicly available at \href{https://pdr.obspm.fr/pdr_download.html}{pdr.obspm.fr}), with extinction properties typical of dense molecular gas as used in \cite{habart2023pdrs4all} for the Orion Bar (Rv = 5.5, NH/E(B-V) = 1.6$\times 10^{22}$~\cmsq), an extinction curve using the fit of \cite{Fitzpatrick_Massa1986} for HD38087, and a power-law grain size distribution extending from 3~nm to 0.3 microns (the model grid is publicly available at \href{https://app.ism.obspm.fr/ismdb/}{app.ism.obspm.fr/ismdb}). \cite{BerneO_2024_science} have shown for a similar analysis that the use of larger grains does not affect strongly the results. 
We account for geometrical effects (beam dilution, inclination of PDR surface with respect to the line of sight and to the direction of irradiation) by adding a single geometrical scaling parameter alpha applied to all \ch{H2} line intensities, as all detected \ch{H2} lines are expected to be optically thin.
We restrict our search of a best fitting model to $G_0$ values between 2$\times$10$^4$ and 8$\times$10$^4$ to account for the previous constraint provided by \cite{haworth2023vlt}, and to values of the geometrical scaling factor alpha between 0.1 and 10, as more extreme values would correspond to unlikely geometries. The range of gas densities explored goes from 1$\times10^1$ to 1$\times10^{10}$ cm-3. The best fitting model has a gas density n$_H$ = 3.1$\times$10$^7$ cm-3, a $G_0$ = 2$\times10^4$ and a geometrical scaling factor alpha = 0.1.}

\subsection*{Photoevaporation mass loss rate} 

Following \cite{BerneO_science}, we compute the photoevaporation mass-loss rate using $\dot{M}_{\rm PE}=S_{\rm PE}~\mu~m_{\rm H}~n_{\rm H}~c_{s}$, 
with $c_{\rm S} \equiv \sqrt{ \frac{7/5 k_{\rm B} T_{\rm gas} }{\mu~m_{\rm H}}}$=3.3~km~s$^{-1}$,
where $k_{\rm B}$ is the Boltzmann constant, $m_{\rm H}$ is the mass of hydrogen and
$\mu$ is the ratio of total mass over hydrogen mass of interstellar gas
($\mu =1.4$ \cite{draine2011}). {The emitting photoevaporation surface area }$S_{\rm PE}$ is computed using {an ellipsoid} \cite{BerneO_science}:
\begin{equation}
    S_{\rm PE}=2 \pi r_{d}^2 + \frac{\pi (E_d/2)^2}{e} \ln{\frac{1+e}{1-e}},
    \label{eq_spheroid}
\end{equation}
where the ellipticity $e=\sqrt{1-(E_d/2) ^2/r_{d}^2}$.
With the parameters reported in Table~\ref{table:system_parameters}, this yields $\dot{M}_{\rm PE}=7.9\times10^{-7}$\mpy.

\subsection*{\ch{H2O}, \ch{CO}, \ch{CH3+} absorption and emission modeling}

\subsubsection*{Analytical models}

To model \ch{H2O}, \ch{CO}, \ch{CH3+} features observed in the data, we utilize slab models. To fit the flux of d203-504 in spectral regions with molecular emission we use:
\begin{equation}
F_{\rm mod}^{em}(\lambda)=S \times B(\lambda, T_{\rm ex}) (1-e^{- \tau^{\rm em}({\lambda, T_{\rm ex}})})+F_{\rm NIR}(\lambda)+F_{\rm MIR}(\lambda),
    \label{eq:model_em}
\end{equation}
where $S$ is the area of the emitting surface, $B_{\lambda, T}$ the Planck function 
and $\tau^{\rm em}(\lambda)$ is the molecular emission optical depth. 
$F_{\star}$ and $F_{\rm NIR}$, $F_{\rm MIR}$ are defined in Eqs.~\ref{f_star}, \ref{f_nir}, \ref{f_mir}, respectively.  
For the absorption, we use:
\begin{equation}
        F_{\rm mod}^{\rm abs}(\lambda)= (F_{\star}(\lambda) + F_{\rm NIR}(\lambda))\times e^{- \tau^{\rm abs}({\lambda, T_{\rm ex}})} + F_{\rm MIR}(\lambda),       
    \label{eq:model_abs}
\end{equation}
where $\tau^{\rm abs}(\lambda, T_{\rm ex})$ is the molecular absorption optical depth and $T_{\rm ex}$ is the excitation temperature. 
{In this case, since the molecules are seen in absorption, the fitting procedure is thus independent of the emitting surface area.} 
We note that only the near infrared component of dust emission ($F_{\rm NIR}$) is absorbed in this model. This is motivated by tests on \ch{H2O} absorption: the best-fit models for the $v=1 \rightarrow 0, \nu_1, \nu_2$ stretching modes show excess absorption in the $\nu_2$ rovibrational band in the MIRI spectrum when both $F_{\rm NIR}$ and $F_{\rm MIR}$ contribute to the background for absorption (Figure~\ref{fig:water}). This confirms that water absorbs only the near infrared emission from the inner hot disk. 
The optical depths (emission and absorption) are defined by $\tau(\lambda, T_{\rm ex})=\sigma({\lambda}, T_{\rm ex}) \times N$ with $\sigma(\lambda,T_{\rm ex})$ the molecular cross section for an excitation temperature $T_{\rm ex}$ , and $N$ the species' column density.
For \ch{H2O} and \ch{CO} we use the cross sections from the {\sc ExoMol} database \citep{Exomol_ref}.
For \ch{CH3+} we rely on the spectroscopic constants provided by \cite{changala2023astronomical}. We use the 
{\sc pgopher} software \citep{western2017:PGOPHER} to  directly compute the cross section assuming a Boltzmann 
distribution to calculate the population in the upper state and no population in the lower state. 
For each molecule, $\sigma({\lambda}, T_{\rm ex})$ is convolved with a Gaussian line profile whose full width half maximum (FWHM) 
is the sum of the instrumental line spread function (derived using a Gaussian fit on narrow emission lines, {which typically gives a FWHM of $\approx120$~\kms.}) and the line Doppler broadening $\Delta V$ expressed in velocity space.  
In addition, the cross section is shifted in wavelength in order to account for the Doppler shift of the emitting/absorbing 
gas with respect to its vacuum rest wavelength. This shift $V$ is given here with respect to the systemic velocity $V_{\rm sys}=25$~\kms 
(see Sect. 2.1 in \cite{haworth2023vlt}). 
{$V$ and $\Delta V$ are results of the fitting procedure and are thus not expected to have preferred values.}
All models are optically thin, thus the surface $S$ in equation \ref{eq:model_em} and associated 
column densities are degenerate and cannot be constrained independently. Thus, for the emissions, the column densities are computed assuming 
a realistic surface. For \ch{CH3+} we consider extended emission from the photodissociation region at the disk surface formed by external far-UV photons from { $\theta^2$ Ori A}. This case is similar to the emission observed in d203-506 \cite{BerneO_science}. Thus, we use $S= \pi r_{\rm d}^2$. 
For the \ch{CO} emission, since the origin is not precisely known, we use this same value for $S$. 

\subsubsection*{Fitting procedure}

{ The models of Eqs.~\ref{eq:model_em} for the emission and \ref{eq:model_abs} for the absorption, are fitted to the observations. To do this, we compute a grid of models $F_{\rm mod} (\lambda)$, with varying values of $T_{\rm ex}$ and $N$. We then compute the residuals $R=\sqrt{F_{\rm obs}^2-F_{\rm mod}^2}$ as a function of these two parameters.
Figure~\ref{fig:residuals} shows an example of the variation of the residuals as a function of $T_{\rm ex}$ and $N$ for the case of 
\ch{H2O} and \ch{CO} absorptions.  The minimum of the residuals provides the best fit model. The uncertainties on $T_{\rm ex}$ and $N$ correspond the values for these parameters found 10\% above the minimum residual.
The resulting values of $T_{\rm ex}$ and $N$  for all species are presented in Table~\ref{table:model_properties}.
Hereafter we describe in more details how the fits were performed for specific spectral signatures. 

For {the} CO emission {fits,} the model is fit in spectral regions where there is no CO absorption and only emission (Figure~\ref{fig:co_em}). This is made possible by the fact that emission occurs at higher excitation temperatures, and thus, the edges of the P and R branches of the CO ro-vibrational spectrum are more extended than the lower temperature absorption.
This yields $T_{\rm CO, em}=2050$~K with a column density of $N_{\rm CO, em}=4.4\times10^{11}$~\cmsq. CO emission lines are the only ones that require $ V \neq 0$ and $\Delta V \neq 0$ (see Table.~\ref{table:model_properties}). The fact that $ V \neq 0$ 
suggests an outflow origin for this emission. 

For \ch{CO} absorption, the model is fit in the NIRSpec range (4.38$-$5.02~\um). We first subtract the CO emission model (see above) to ease the fitting of the absorption.  
For \ch{H2O} absorption, the model is fit in the NIRSpec range (2.45$-$2.91~\um). 
The best fit models for \ch{H2O} and \ch{CO} absorption obtained using the procedure relying on residuals (see above) 
are shown in Figure~\ref{fig:models}. In this figure, we present the
model optical depth ($\tau^{\rm abs} (\lambda, T_{\rm ex})$) and the observationally derived optical depths, that is :
$\tau^{\rm obs} = log (F_{\rm obs}(\lambda) / F_{\rm NIR}(\lambda))$, where $F_{\rm obs}(\lambda)$ is the observed spectrum. This best fit corresponds to $T_{\rm H2O}=$850$\pm250$~K and $T_{\rm CO}=$1150$\pm$350~K with column densities of $N_{\rm H2O}=9.1\pm1.4 \times 10^{17}$cm$^{-2}$ and $N_{\rm CO}=0.8\pm0.1 \times 10^{18}$cm$^{-2}$. 

For \ch{CH3+} emission, the model is fit in the MIRI range (7.0$-$7.35~\um) using the same approach as described above. The best fit is shown in Figure~\ref{fig:pah_ch3p} and gives  $T=1000\pm100$~K and $N=4.0\times 10^{13}$~\cmsq.

\subsubsection*{Modeling of water absorption in the MIRI range}

{ Using the column density and temperature for water derived from the fit of the NIRSpec data (see above), we compute the predicted spectrum in the MIRI range. We consider two cases:
\begin{itemize}
    \item case 1: the absorption occurs against the stellar and the near infrared emission from dust in the inner disk (as described by Eq.~\ref{eq:model_abs});
    \item case 2: the absorption occurs against the whole infrared continuum emission, that is :
        \begin{equation}
        F_{\rm mod}(\lambda)= (F_{\star}(\lambda) + F_{\rm NIR}(\lambda)+F_{\rm MIR}(\lambda))\times e^{- \tau^{\rm abs}({\lambda, T_{\rm ex}})}.       
        \label{eq:model_abs_miri}
\end{equation} 
\end{itemize}

The resulting models for these two cases are shown in Figure~\ref{fig:water} in the MIRI range, near 6.7 \um. It can be seen in this figure that the model for the case 2 produces absorption features which are much stronger than observed, while case 1 agrees better with observations. This provides an additional confirmation that water absorption originates from the the inner disk. }

\section*{Data Availability}

The JWST data presented in this paper is publicly available through the MAST online archive (\url{http://mast.stsci.edu}) using the PID 1288. 

\section*{Code Availability}
The JWST pipeline used to produce the final data products presented in this article is available at
\href{https://github.com/spacetelescope/jwst}{https://github.com/spacetelescope/jwst}.
The code used in this study is available from the corresponding author on reasonable request.

\section*{Acknowledgments}
This work is based [in part] on observations made with the NASA/ESA/CSA James Webb Space Telescope. The data were obtained from the Mikulski Archive for Space Telescopes at the Space Telescope Science Institute, which is operated by the Association of Universities for Research in Astronomy, Inc., under NASA contract NAS 5-03127 for JWST. These observations are associated with programs \#1288.
IS, OB are funded by the Centre National d'Etudes Spatiales (CNES) through the APR program. 
This research received funding from the program ANR-22-EXOR-0001 Origins of the Institut National des Sciences de l’Univers, CNRS. 
We thank R. Le Gal for her comments on the manuscript.
TJH acknowledges funding from a Royal Society Dorothy Hodgkin Fellowship and UKRI guaranteed funding for a Horizon Europe ERC consolidator grant (EP/Y024710/1).
JRG thanks the Spanish MCINN for funding support under grant PID2023-146667NB-I00.
This project is co-funded by the European Union (ERC, SUL4LIFE, grant agreement No 101096293). AF also thanks project PID2022-137980NB-I00 funded by the Spanish
Ministry of Science and Innovation/State Agency of Research MCIN/AEI/10.13039/501100011033 and by “ERDF A way of making Europe”.
TO acknowledges the support by the Japan Society for the Promotion of Science (JSPS)
KAKENHI Grant Number JP24K07087.
CB is grateful for an appointment at NASA Ames Research Center through the San José State University Research Foundation (80NSSC22M0107). 
{ We thank the referees for their comments and suggestions, which helped to improve the clarity of the manuscript.}
\section*{Author Contribution Statement}

I.S. and O.B. wrote the manuscript with significant input from E.B..
I.S. did the line analysis with support from O.B..
I.S., A.C., R.C., A.S., B.T., F.A., D.V.P. reduced the data. 
E.H., E.P., and O.B. planned and co-led the ERS PDRs4All program. 
O.B., E.B, F.,A., P.A., E.A.B, C.B, J.C., G.A.L.C., E.D., A.F., J.R.G., E.H., T.J.H., C.J., F.L.P., T.O., E.P., M.R., A.G.G.M.T. and M.Z. contributed to the observing program with JWST. 
All authors participated in either the development and testing of the MIRI-MRS or NIRSpec instruments and their data reduction, in the discussion of the results, and/or commented on the manuscript.

\section*{Competing Interest Statement}

The authors declare no competing interests.


\section*{Tables}
\setcounter{table}{0}
\renewcommand{\thetable}{Extended Data~\arabic{table}}

\newpage\setcounter{page}{1}

\begin{table}
\centering
\caption{Detected emission lines and their intensities}
\label{table:line_intensities}
\begin{tabular}{lcr}
\hline
Species    &  Wavelength & Intensity \\
(1)   &  (2)   & (3)\\ 
\hline
\HI         & 1.0052 & 7.3092 $\times 10^{-4}$  \\
\ch{H2}     & 1.0054 & 3.3239 $\times 10^{-4}$ \\
$[\SII]$    & 1.0289 & 1.6407 $\times 10^{-4}$ \\
$[\SII]$    & 1.0323 & 2.2115 $\times 10^{-4}$ \\
$[\SII]$    & 1.0339 & 1.5751 $\times 10^{-4}$ \\
$[\SII]$    & 1.0373 & 7.2280 $\times 10^{-5}$  \\
\HeI        & 1.0832 & 60.299 $\times 10^{-4}$ \\
\ch{H2}     & 1.0838 & 28.379 $\times 10^{-4}$ \\
\HI         & 1.0941 & 17.987 $\times 10^{-4}$ \\
\HeI        & 1.2788 & 8.2167 $\times 10^{-5}$  \\
\HI         & 1.2821 & 29.607 $\times 10^{-4}$ \\
\ch{H2}     & 1.2823 & 13.536 $\times 10^{-4}$ \\
\OI         & 1.3168 & 6.8693 $\times 10^{-5}$  \\
\HeI        & 1.4486 & 5.1291 $\times 10^{-5}$ \\
\ch{H2}     & 1.4591 & 6.0641 $\times 10^{-5}$ \\
\HI         & 1.5345 & 6.6678 $\times 10^{-5}$  \\
\HI         & 1.5443 & 5.1240 $\times 10^{-5}$  \\
\HI         & 1.5560 & 6.8D6 $\times 10^{-5}$  \\
\HI         & 1.5704 & 8.2517 $\times 10^{-5}$  \\
\HI         & 1.5884 & 8.8201 $\times 10^{-5}$  \\
\HI         & 1.6113 & 1.6521 $\times 10^{-4}$  \\
\HI         & 1.6411 & 1.6898 $\times 10^{-4}$     \\
$[\FeII]$   & 1.6440 & 7.6064 $\times 10^{-5}$  \\
\HI         & 1.6811 & 2.0592 $\times 10^{-4}$ \\
\HeI        & 1.7006 & 5.6076 $\times 10^{-5}$  \\
\ch{H2}     & 1.7364 & 8.3027 $\times 10^{-5}$  \\
\HI         & 1.7366 & 1.0473 $\times 10^{-4}$ \\
\ch{H2}     & 1.7369 & 1.0487 $\times 10^{-4}$ \\
\HI         & 1.8179 & 2.8084 $\times 10^{-4}$  \\
\ch{H2}     & 1.8182 & 7.3514 $\times 10^{-5}$  \\
\HeI        & 1.8690 & 2.6438 $\times 10^{-4}$  \\
\HeI        & 1.8702 & 1.0561 $\times 10^{-4}$   \\
\HI         & 1.8756 & 89.846 $\times 10^{-4}$   \\
\HI         & 1.9450 & 4.9110 $\times 10^{-4}$  \\
\HeI        & 2.0586 & 5.9400 $\times 10^{-4}$  \\
\ch{H2}     & 2.1218 & 7.1025 $\times 10^{-5}$  \\
\HI         & 2.1661 & 8.1658 $\times 10^{-4}$  \\
\ch{H2}     & 2.3397 & 8.9336 $\times 10^{-5}$  \\
\ch{H2}     & 2.4065 & 1.5008 $\times 10^{-4}$ \\
\HeI        & 2.4127 & 5.8390 $\times 10^{-5}$  \\
\ch{H2}     & 2.4135 & 9.0322 $\times 10^{-5}$  \\
\HI         & 2.4163 & 9.9279 $\times 10^{-5}$  \\
\ch{H2}     & 2.4208 & 5.7382 $\times 10^{-5}$ \\
\ch{H2}     & 2.4296 & 5.4590 $\times 10^{-5}$ \\
\hline
\end{tabular}
{
(1) Species name;
(2) Emission line wavelength (\um), from the PDRs4All line list;
(3) Line intensity (erg~s$^{-1}$~cm$^{-2}$~sr$^{-1}$).
}
\end{table}
\setcounter{page}{1}
\begin{table}
\centering
\caption{Table Extended 
1, continued.}
\label{table:line_intensities_2}
\begin{tabular}{lcr}
\hline
Specie    &  Wavelength& Intensity \\
(1)   &  (2)   & (3) \\
\hline
\HI         & 2.4313 & 1.1551 $\times 10^{-4}$ \\
\ch{H2}     & 2.4354 & 5.5206 $\times 10^{-5}$   \\
\ch{H2}     & 2.4374 & 7.3335 $\times 10^{-5}$  \\
\ch{H2}     & 2.4428 & 6.4722 $\times 10^{-5}$  \\
\HI         & 2.4490 & 5.2216 $\times 10^{-5}$  \\
\ch{H2}     & 2.4547 & 5.9707 $\times 10^{-5}$  \\
\HI         & 2.6126 & 1.0249 $\times 10^{-4}$ \\
\HeI        & 2.6241 & 1.13415$\times 10^{-4}$  \\
\HeI        & 2.6259 & 14.625 $\times 10^{-4}$ \\
\ch{H2}     & 2.6268 & 1.3713 $\times 10^{-4}$ \\
\HI         & 2.6751 & 9.0793 $\times 10^{-5}$  \\
\HI         & 2.7582 & 1.5352 $\times 10^{-4}$ \\
\HI         & 2.8729 & 1.5189 $\times 10^{-4}$ \\
\HI         & 3.0392 & 1.7567 $\times 10^{-4}$  \\
\HI         & 3.2969 & 2.3034 $\times 10^{-4}$ \\
\ch{H2}     & 3.2981 & 7.0150 $\times 10^{-5}$  \\
\HI         & 3.7405 & 3.7225 $\times 10^{-4}$  \\
\HeI        & 4.0490 & 9.8952 $\times 10^{-5}$  \\
\HI         & 4.0522 & 21.880 $\times 10^{-4}$  \\
\ch{H2}     & 4.0540 & 3.0313 $\times 10^{-4}$  \\
$[\FeII]$   & 4.0763 & 1.1952 $\times 10^{-4}$ \\
\ch{H2}     & 4.0806 & 9.0220 $\times 10^{-5}$  \\
$[\FeII]$   & 4.0819 & 1.0560 $\times 10^{-4}$ \\
$[\FeII]$   & 4.1149 & 5.2802 $\times 10^{-5}$ \\
\HeI        & 4.1227 & 5.4020 $\times 10^{-5}$ \\
\HI         & 4.3764 & 7.5432 $\times 10^{-5}$  \\
\HI         & 4.6537 & 5.4738 $\times 10^{-4}$    \\
\ch{H2}     & 4.6557 & 9.6589 $\times 10^{-5}$  \\
\HI         & 4.6725 & 1.2686 $\times 10^{-4}$    \\
\HI         & 5.0913 & 5.9017 $\times 10^{-5}$  \\
\HI         & 5.1286 & 1.5268 $\times 10^{-4}$ \\
$[\FeII]$   & 5.3401 & 7.7037 $\times 10^{-5}$  \\
\ch{H2}     & 5.5111 & 7.8753 $\times 10^{-5}$   \\
\HI         & 5.9082 & 1.8906 $\times 10^{-4}$ \\
\ch{H2}     & 6.9095 & 1.8656 $\times 10^{-4}$  \\
$[\ArII]$   & 6.9852 & 4.7226 $\times 10^{-4}$  \\
\HI         & 7.4598 & 7.1026 $\times 10^{-4}$  \\
\HI         & 7.5024 & 1.7393 $\times 10^{-4}$ \\
\HI         & 7.5081 & 7.2068 $\times 10^{-5}$  \\
\ch{H2}     & 8.0250 & 8.7383 $\times 10^{-5}$  \\
\HI         & 8.7600 & 5.6647 $\times 10^{-5}$ \\
$[\ArIII]$  & 8.9913 & 38.084 $\times 10^{-4}$ \\
\ch{H2}     & 9.6649 & 1.1071 $\times 10^{-4}$ \\
\HI         & 10.503 & 5.5302 $\times 10^{-5}$  \\
$[\SIV]$    & 10.510 & 4.2879 $\times 10^{-4}$  \\
$[\NiII]$   & 10.682 & 6.1122 $\times 10^{-5}$  \\
\HeI        & 10.799 & 5.6502 $\times 10^{-5}$  \\
\HI         & 10.803 & 5.7912 $\times 10^{-5}$  \\
\HI         & 11.308 & 9.5396 $\times 10^{-5}$  \\
\HI         & 12.371 & 1.7725 $\times 10^{-4}$ \\
$[\NeII]$   & 12.813 & 134.94 $\times 10^{-4}$ \\
$[\NeIII]$  & 15.555 & 1.5734 $\times 10^{-4}$ \\
$[\SIII]$   & 18.713 & 22.196 $\times 10^{-4}$ \\
$[\ArIII]$  & 21.829 & 6.2053 $\times 10^{-5}$  \\
$[\FeIII]$  & 22.925 & 7.1886 $\times 10^{-5}$  \\
\hline
\end{tabular}
\end{table}

\begin{table}
\centering
\caption{\ch{H2} lines detection properties}
\label{table:h2_em}
\begin{tabular}{lcr}
\hline
Transition    &  Wavelength &  Intensity ($\times 10^{-5}$) \\
(1)   &  (2)   & (3) \\ 
\hline
$v=0 \rightarrow 0$ ($J=3 \rightarrow 1$) & 17.034 & 4.865 \\
$v=0 \rightarrow 0$ ($J=4 \rightarrow 2$) & 12.278 & 3.383 \\
$v=0 \rightarrow 0$ ($J=5 \rightarrow 3$) & 9.6649 & 11.07 \\
$v=0 \rightarrow 0$ ($J=6 \rightarrow 4$) & 8.0250 & 8.738 \\
$v=0 \rightarrow 0$ ($J=7 \rightarrow 5$) & 6.9095 & 18.65 \\
$v=0 \rightarrow 0$ ($J=8 \rightarrow 6$) & 6.1085 & 2.978 \\
$v=0 \rightarrow 0$ ($J=9 \rightarrow 7$) & 5.5111 & 7.875 \\
$v=0 \rightarrow 0$ ($J=10 \rightarrow 8$) & 5.0531 & 2.414 \\
$v=1 \rightarrow 0 $ ($J=2 \rightarrow 4$) & 3.0038 & 2.640 \\
$v=1 \rightarrow 0 $ ($J=2 \rightarrow 0$) & 2.2232 & 2.917 \\
$v=1 \rightarrow 0 $ ($J=3 \rightarrow 5$) & 3.2349 & 2.959 \\
$v=1 \rightarrow 0 $ ($J=3 \rightarrow 1$) & 2.1218 & 7.102 \\
$v=1 \rightarrow 0 $ ($J=4 \rightarrow 4$) & 2.4374 & 7.333 \\
$v=1 \rightarrow 0 $ ($J=4 \rightarrow 2$) & 2.0337 & 3.184 \\
$v=1 \rightarrow 0 $ ($J=5 \rightarrow 5$) & 2.4547 & 5.970 \\
$v=1 \rightarrow 0 $ ($J=5 \rightarrow 3$) & 1.9575 & 6.180 \\
$v=1 \rightarrow 0 $ ($J=7 \rightarrow 5$) & 1.8357 & 2.008 \\
\hline
\end{tabular}
{
(1) \ch{H2} transition label;
(2) Emission line wavelength (\um), from \cite{roueffH2};
(3) Line intensity (erg~s$^{-1}$~cm$^{-2}$~sr$^{-1}$).
}
\end{table}

\clearpage
\newpage

\begin{table*}
\centering
\caption{Derived Physical Properties of Considered Species.}
\label{table:model_properties}
\begin{tabular}{lccccccr}
\hline
ID  & Detection &  $T_{\rm ex}$ & $N$  & $\Delta V$ & $V$\\
(1)     &  (2)      & (3)    &    (4)   &       (5)       &    (6) \\
\hline
\ch{H2O}  & abs & 850$\pm$ 250  & 9.1$\pm1.4\times 10^{17}$   & 0  & 0\\
\ch{CO}   & abs & 1150$\pm$ 350 & 0.8$\pm 0.1\times 10^{18}$   & 0  & 0\\
\ch{CH3+} & em  & 1000$\pm$ 100 &    4.0$\pm 0.8\times 10^{13}$ & 0  & 0\\
\ch{CO}   & em  & 2050$\pm$ 100 &  4.4$\pm 0.9\times 10^{11}$&  80  & -25\\
 \hline
\end{tabular}\\
{ \small \vspace{0.5cm}
(1) Species;
(2) Detected in emission (em) or in absorption (abs);
(3) Excitation temperature (in K);
(4) Column density (cm$^{-2}$), see details in the text on the assumptions for the size of the emitting surface in the case of emission;
(5) Line Doppler broadening (in \kms);
(6) Doppler shift with respect to systemic velocity of d203-504 (25 \kms)
}
\end{table*}

\clearpage
\newpage


\section*{Figure Legends/Captions}

Figure~\ref{spectrum}:{ JWST NIRSpec/IFU and MIRI/MRS combined spectrum of the d203-504 proplyd}. The vertical lines indicate the position of HI (dashed), He I (dotted) and \ch{H2} (continuous) emission lines. Fine structure lines of atoms are labeled directly on the figure.
Three close-up views are displayed in framed insets: water absorption at 2.7~\um\ in blue ({ a)}), CO absorption and emission at 4.7~\um\ in purple ({ b)}) and \ch{CH3+} emission at 7.2~\um\ in orange ({ c)}). Emission bands from PAHs are highlighted in green.\\

Figure~\ref{fig:pah_ch3p}:{ Analysis of PAH, \ch{CH3+} and \ch{H2} emissions in d203-504}.
{ a)} Continuum subtracted PAH spectrum ($F_{\rm PAH}(\lambda)$, see Methods) in arbitrary units, for d203-504 (gray) and the Orion Bar PDR (green). 
{ b)} Continuum subtracted spectrum of d203-504 ($F_{\rm obs}(\lambda) - F_{\rm cont}(\lambda)$, see Methods) in the spectral region of  \ch{CH3+} emission, in dark gray and the best fit model in blue.
{ c)} Comparison between the observed (blue) and modeled (orange) \ch{H2} line intensities using the Meudon PDR model. Error bars represent a total uncertainty of 50\% used in the minimization to obtain the best fit model. The instrumental uncertainties are smaller than the point size and given in Table~\ref{table:model_properties}.
Notations of \ch{H2} transitions on the x-axis are abbreviated. The corresponding quantum levels are given in Table~\ref{table:h2_em}.\\

Figure~\ref{fig:models}:{ Optical depth spectra of water and carbon monoxide absorption in the inner disk of d203-504.} { a)} Water absorption optical depth spectrum derived from the observations ($\tau_{\rm obs}$, see Methods), in black, and best fit model ($\tau_{\rm abs}$) shown in negative, in blue. 
{ b)} Carbon monoxide absorption optical depth spectrum derived from the observations, and best fit model in blue.
The gray areas correspond to regions of contamination by strong atomic lines where the data has been {excluded from the fit}.  
Column densities and temperatures obtained with the best fit models are indicated.\\

Figure~\ref{scheme}:{ Chemical structure in the solar-type d203-504 Proplyd.}
{ a)} Sketch of the inner disk ($\lesssim 1$~au), composed of hot dust grains and water vapor heated by accretion processes. 
This heating sets the position of the snowline, that is, the radius at which the water vapor can freeze-out on dust grains, at $r\sim 0.7$~au (vertical dashed line). 
{ b)}  Sketch of the outer disk ($\gtrsim 1$~au). The surface layers host a photodissociation region with photo-chemistry involving \ch{CH+}, \ch{CH3+}, OH and PAHs driven by the FUV photons from nearby massive stars. 
This PDR triggers mass-loss by photoevaporation.\\

Figure~\ref{fig:fcpah}: Fraction of the available cosmic carbon abundance locked in PAHs. Those fractions are measured at different times of the stellar evolution (from older to more recent, left to right). The Orion Bar and NGC 7023 are photodissociation region in star-forming regions. HST10 is a proplyd in Orion, situated closer to the T{r}apezium cluster {with respect} to d203-504 with $G_0\sim2\times10^5$. Ryugu is a type C asteroid {and is used as an approximation for the Solar System}. See Methods for details on the derivation of PAH abundances presented in this figure.



\newpage


\clearpage
\newpage
\section*{Extended data}
\label{section:extended}
\setcounter{figure}{0}
\renewcommand{\thefigure}{Extended Data~\arabic{figure}}

\begin{figure}[ht!]
  \centering
  \includegraphics[width=13cm]{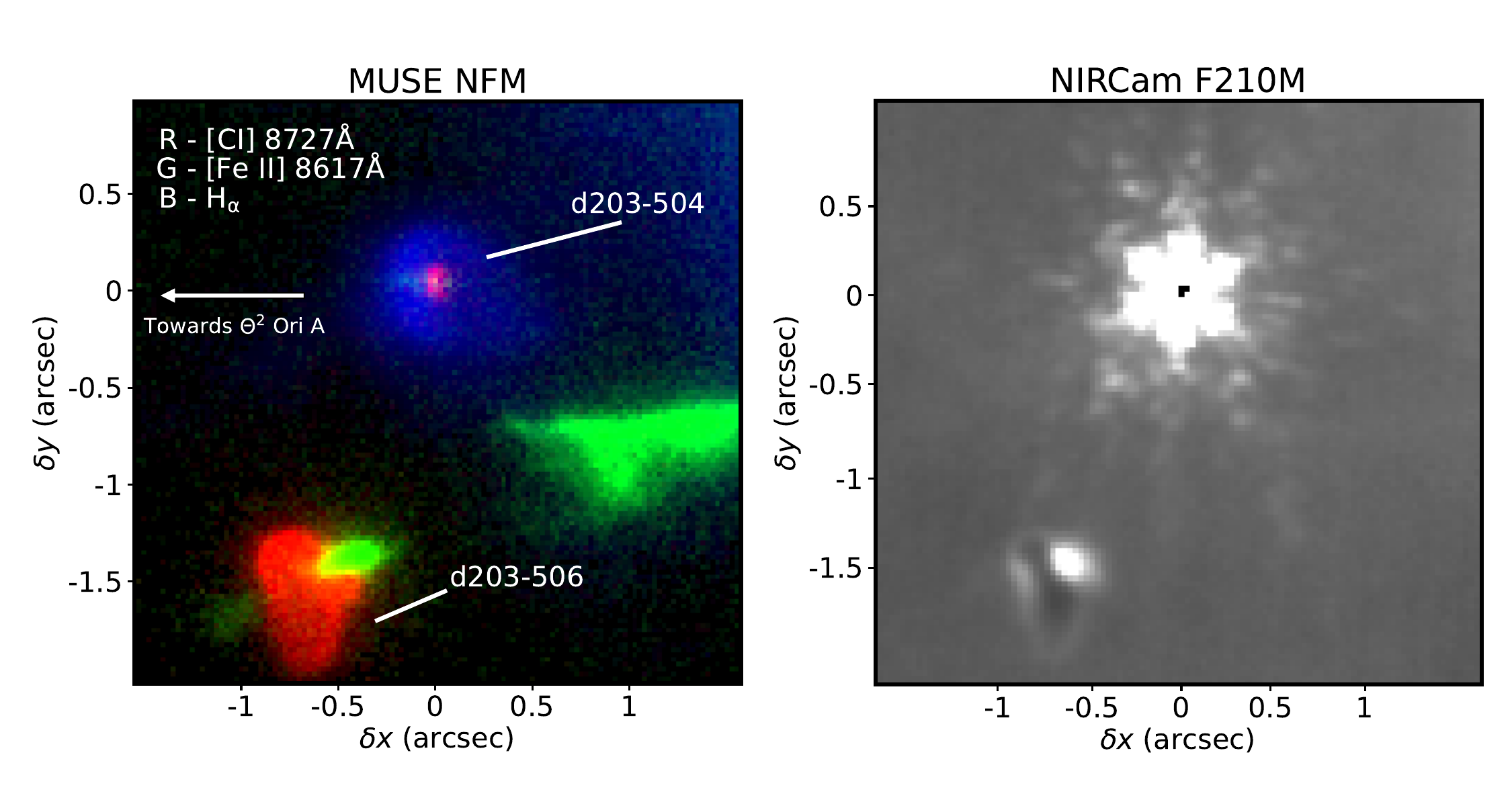}
  \caption{{MUSE and NIRCam images of d203-504 environment.}
  \textit{Left: }MUSE NFM composite image of both d203-506 (bottom left) and d203-504 (top).
  Red: the [\CI]($\lambda 8727\AA$) map showing disk contributions from both d203-506 and d203-504 objects.  
  Green: the [\FeII] ($\lambda 8617\AA$) map corresponding to jets from both objects.
  Blue: the \Ha\ emission map showing the ionization fronts (only d203-504 appears to have one). The glow on the top right corresponds to the Orion Bar ionization front.
  \textit{Right: }JWST NIRCam image of the same system from the F210M filter.
  }
  \label{fig:rgb}
\end{figure}

\begin{figure}[ht!]
  \centering
  \includegraphics[width=8cm]{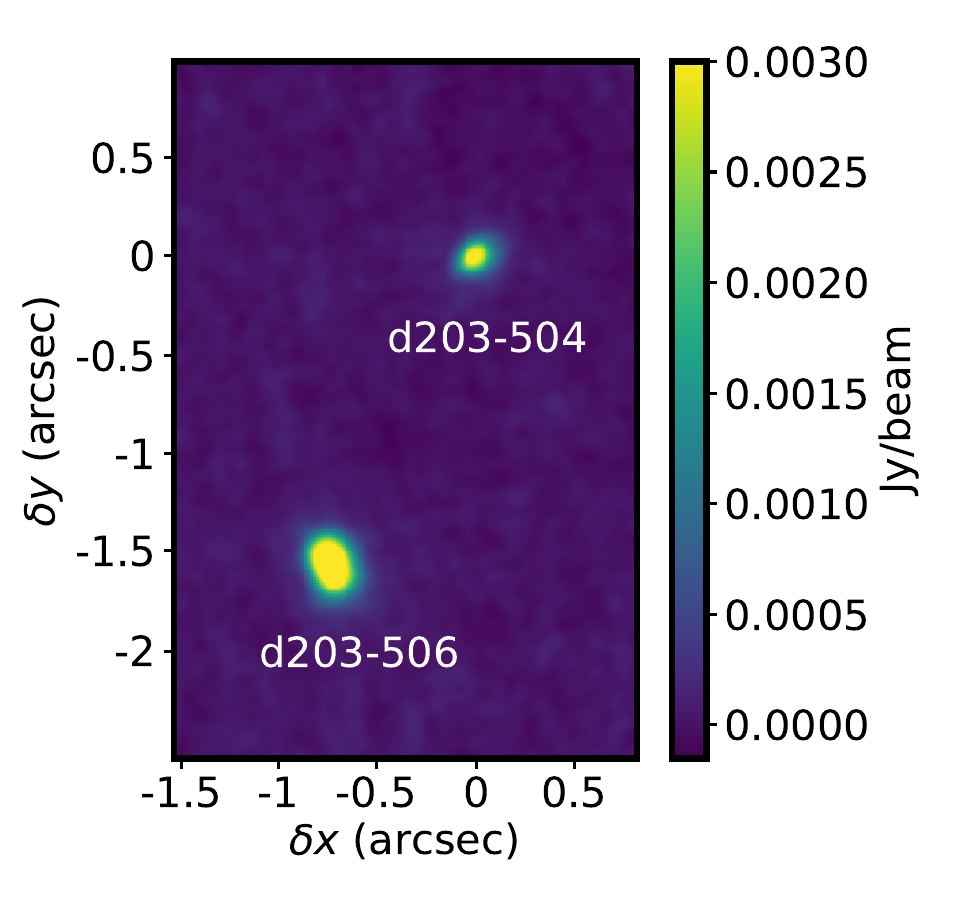}
  \caption{ALMA 344 GHz emission map of d203-506 and d203-504. Values are given in Jy$/$beam and the coordinates are givens as offsets in arcseconds with respect to the position of d203-504  (RA$=$5:35:20.268; DEC$=$-05:25:03.992). 
  }
  \label{fig:alma}
\end{figure}

\begin{figure}[ht!]
  \centering
  \includegraphics[width=13cm]{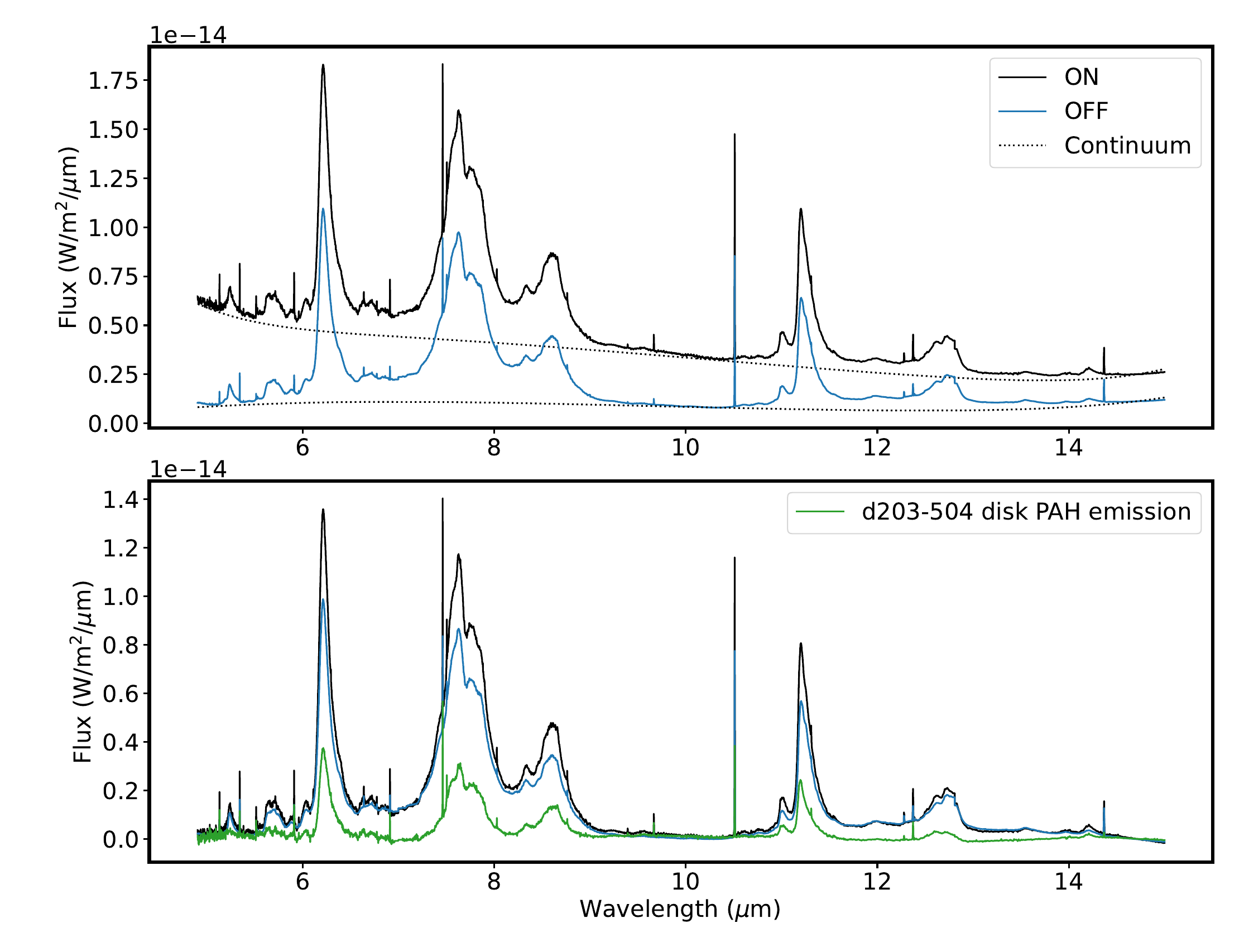}
  \caption{{ON and OFF extracted spectra and PAH emission.}
  {Top panel: }Observed spectrum of d203-504 (the ``ON'' position) in black and the ``OFF'' position in blue. {Fitted continuums are shown in dotted black lines.
  Bottom panel: continuum subtracted spectrum of d203-504, ``ON'' and ``OFF'' positions in black and blue, respectively, and the disk PAH emission in green.}}
  \label{fig:pah_on_vs_off}
\end{figure}

\begin{figure}[h!]
  \centering
  \includegraphics[width=13cm]{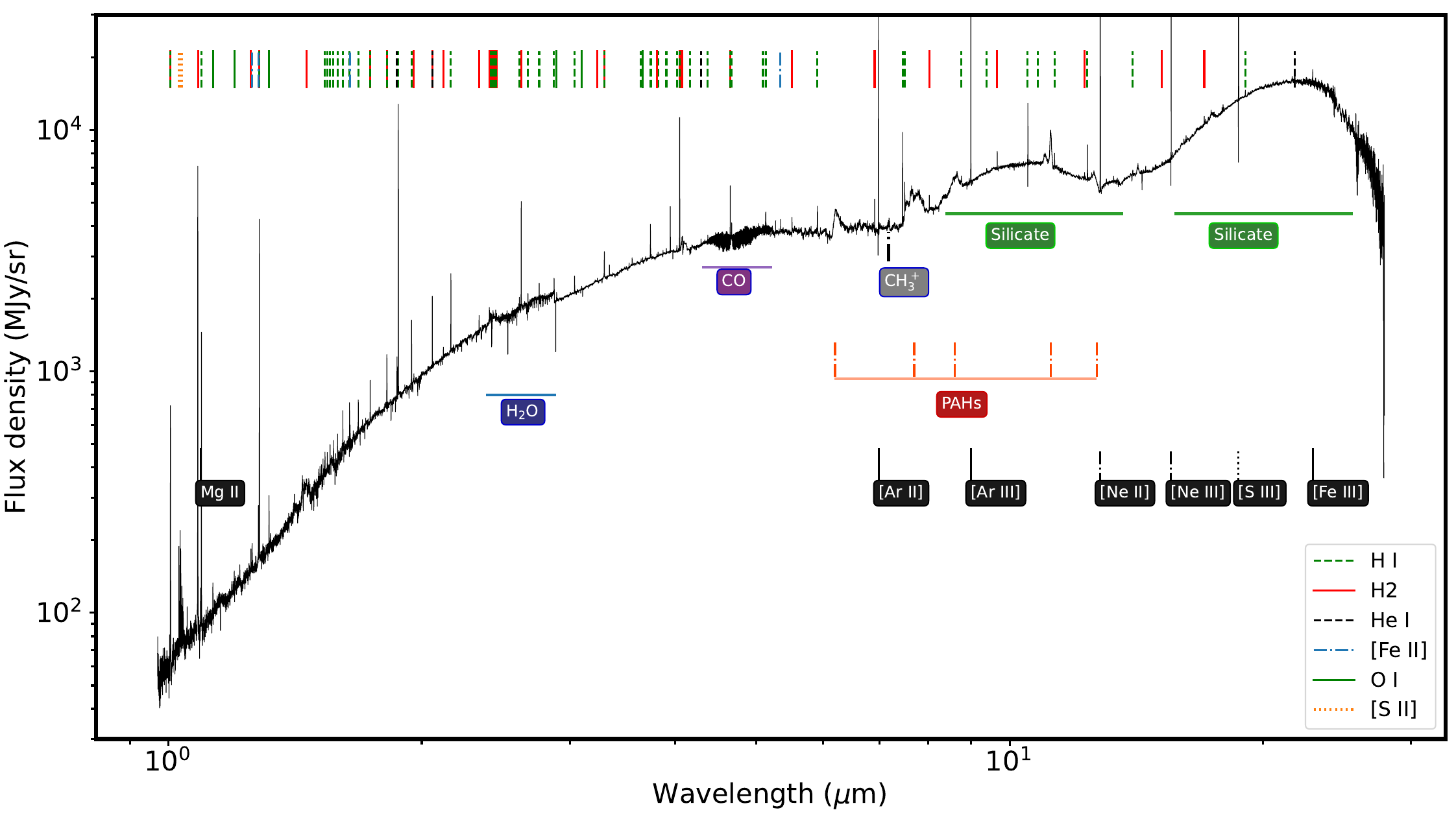}
  \caption{JWST NIRSpec/IFU and MIRI/MRS combined spectrum of the d203-504 proplyd. The vertical lines indicate the position of HI (dashed), He I (dotted) and \ch{H2} (continuous) emission lines. Fine structure lines of atoms are labeled directly on the figure.
This spectrum is in units of MJy/sr to emphasize the silicate emissions highlighted in green.
  }
  \label{fig:spectrum_mjy}
\end{figure}

\begin{figure}[ht!]
  \centering
  \includegraphics[width=13cm]{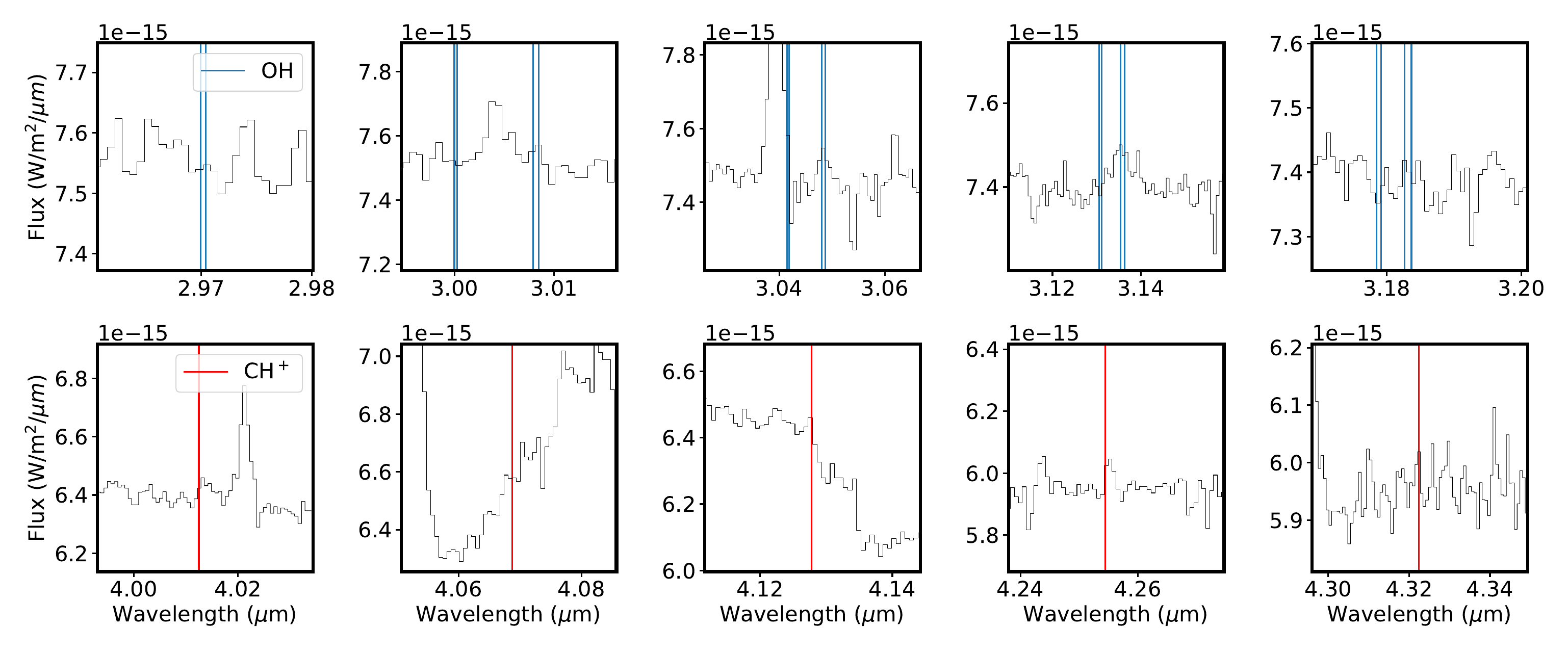}
  \caption{{OH and \ch{CH+} emission.}
  Zooms on d203-504 NIRSpec spectrum, represented in black on both panels. On the upper panels, positions of OH emission lines are shown in blue and on the lower panels, positions of \ch{CH+} emission lines are shown in red. 
  }
  \label{fig:oh_chp}
\end{figure}

\begin{figure}[ht!]
  \centering
  \includegraphics[width=13cm]{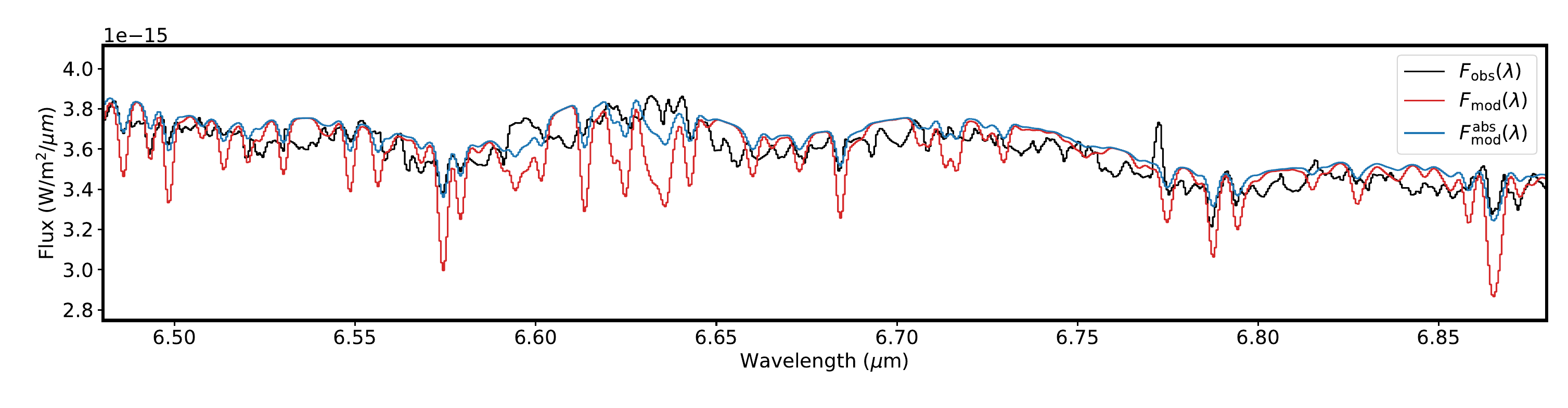}
  \caption{{MIRI water absorption models.}
  Spectroscopic models ($F_{\rm mod}(\lambda)$) of d203-504 using Eq.~\ref{eq:model_abs} (blue) and Eq.~\ref{eq:model_abs_miri} (red).  Observed spectrum ($F_{\rm obs}(\lambda)$) is in black.
  }
  \label{fig:water}
\end{figure}

\clearpage
\newpage
\setcounter{figure}{0}

\renewcommand{\thefigure}{Supplementary~\arabic{figure}}

\newpage

\begin{figure}[ht!]
  \centering
  \includegraphics[width=13cm]{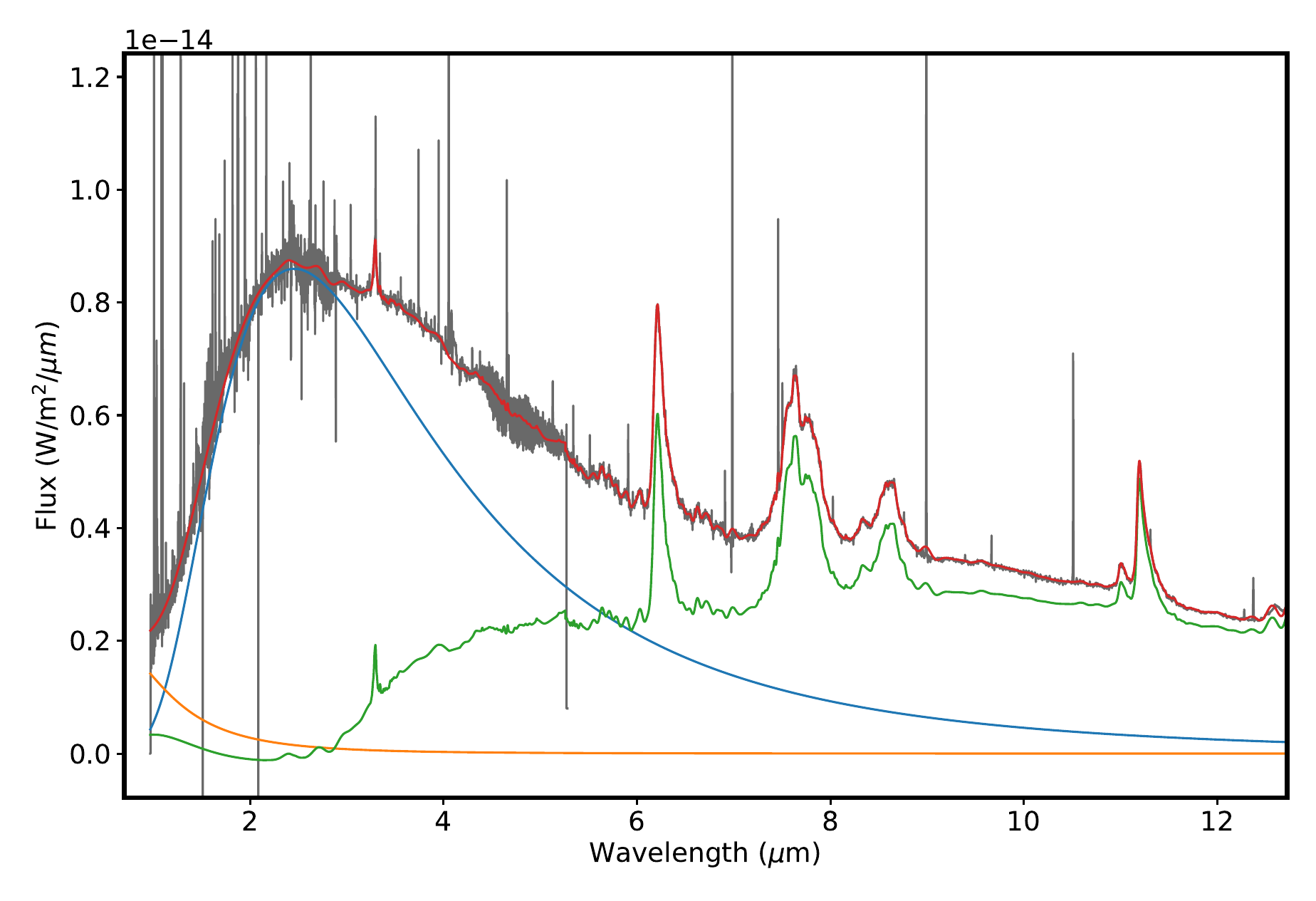}
  \caption{{Modelled continuum for d203-504 observed spectrum.}
  Observed spectrum of d203-504 ($F_{\rm obs}(\lambda)$, gray). Modelled continuum and PAH emission ($F_{\rm cont}(\lambda)$, red), which is the sum of  the stellar $F_{\star}(\lambda)$ (orange), $F_{\rm NIR}(\lambda)$ (blue) and 
  $F_{\rm MIR}(\lambda)$ (green).}
  \label{fig:bb}
\end{figure}

\begin{figure}[ht!]
  \centering
  \includegraphics[width=12cm]{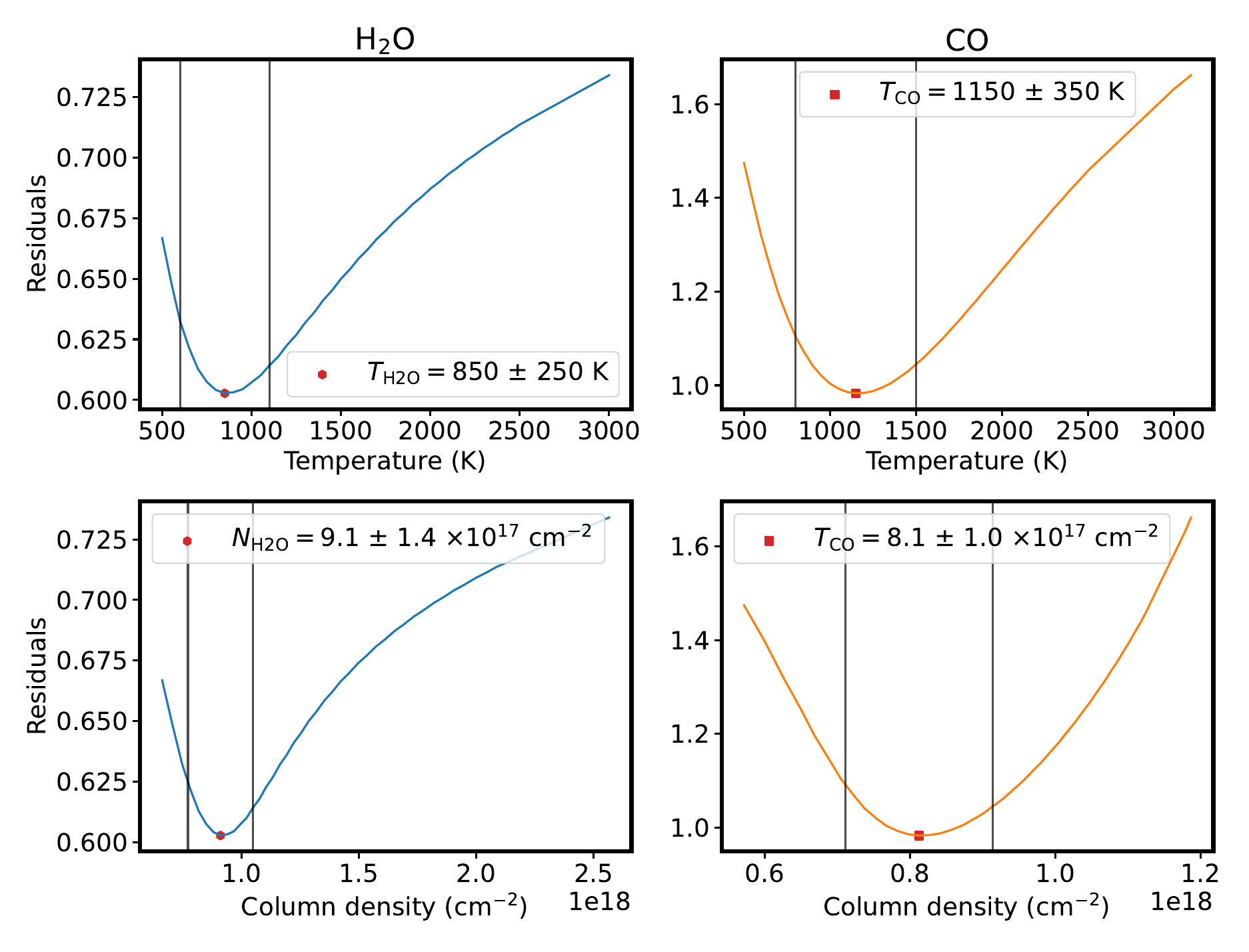}
  \caption{Residuals (root mean square error) of the fits of \ch{H2O} and \ch{CO} models to the observed spectra. The top row represents the temperature residuals for \ch{H2O} (left) and CO (right). The bottom row shows the residuals for column densities of \ch{H2O} (left) and CO (right). Best models are indicated with the red point. {vertical lines indicate the 10\% uncertainties of each best model.}}
  \label{fig:residuals}
\end{figure}

\begin{figure}[ht!]
  \centering
  \includegraphics[width=12cm]{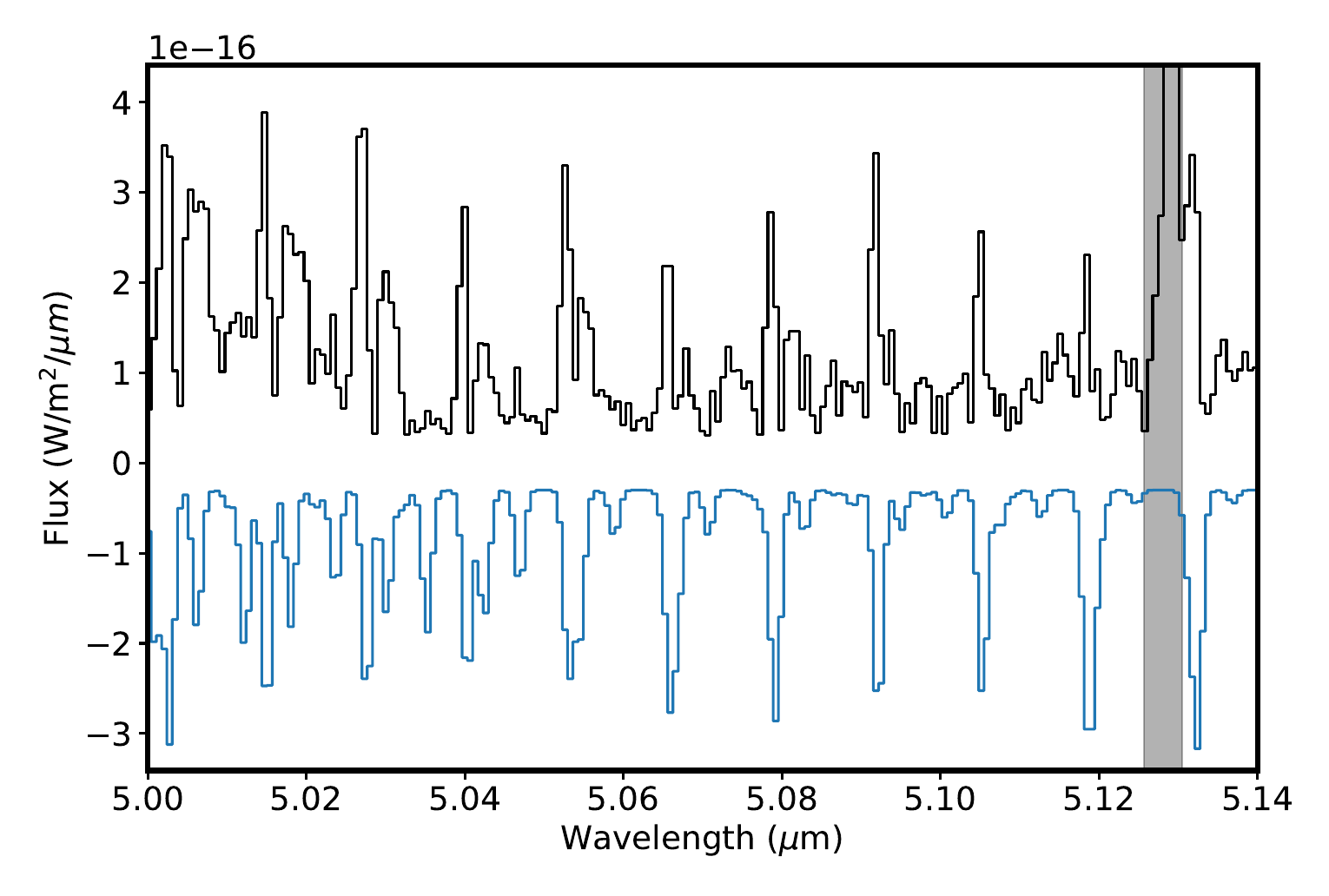}
  \caption{ Continuum subtracted spectrum of d203-504 in the spectral region of CO emission. Data are in black and the inverted model in blue.}
  \label{fig:co_em}
\end{figure}

\end{document}